\documentclass[prodmode,acmtoit]{acmsmall}

\usepackage{booktabs}     
\usepackage{graphicx}
\usepackage{amsmath,amssymb}
\usepackage{subfigure}
\usepackage{fancyvrb}
\usepackage{multirow}
\usepackage{cite}
\usepackage{url}
\usepackage{times}
\usepackage{color}
\usepackage[bf,tableposition=top]{caption}     
\usepackage[bookmarks, pdftex, colorlinks=false]{hyperref}     
\usepackage[square,numbers]{natbib}     
\usepackage{siunitx}	

\newcommand{\ignore}[1]{}
\newcommand{\no}[1]{}
\newcommand{\tc}[0]{\textcolor{black}}

\begin{document}


\markboth{N. Kourtellis et al.}{Enabling Social Applications via Decentralized Social Data Management}

\title{Enabling Social Applications via Decentralized Social Data Management}
\author{NICOLAS KOURTELLIS
\affil{Yahoo Labs, Barcelona, Spain}
JEREMY BLACKBURN
\affil{Department of Computer Science and Engineering, University of South Florida}
CRISTIAN BORCEA
\affil{Department of Computer Science, New Jersey Institute of Technology}	
ADRIANA IAMNITCHI
\affil{Department of Computer Science and Engineering, University of South Florida}
}

\begin{abstract}

An unprecedented information wealth produced by online social networks, further augmented by location/collocation data, is currently fragmented across different proprietary services.
Combined, it can accurately represent the social world and enable novel socially-aware applications.
We present Prometheus, a socially-aware peer-to-peer service that collects social information from multiple sources into a multigraph managed in a decentralized fashion on user-contributed nodes, and exposes it through an interface implementing non-trivial social inferences while complying with user-defined access policies.
Simulations and experiments on PlanetLab with emulated application workloads show the system exhibits good end-to-end response time, low communication overhead and resilience to malicious attacks.

\end{abstract}


\acmformat{Nicolas Kourtellis, Jeremy Blackburn, Cristian Borcea and Adriana Iamnitchi, 2015. Enabling Social Applications via Decentralized Social Data Management. Special Issue on Foundations of Social Computing}

\maketitle

\section{Introduction}\label{sec:intro}

Recent socially-aware applications leverage users' social information to provide features such as filtering restaurant recommendations based on reviews by friends (e.g., Yelp), recommending email recipients or filtering spam based on previous email activity~\cite{kong06collaborative}, and exploiting social incentives for computer resource sharing~\cite{li06f2f,tran08friendstore}.
Social information is also leveraged in conjunction with location and collocation data in mobile applications such as  Loopt and Foursquare, which collect, store, and use sensitive geosocial information.

The state of the art is to collect and manage such information within an application, thus offering social functionalities limited only to the context of the application, as in the examples above, or to expose this information from platforms that specifically collect and manage it, such as online social networks (OSNs).
For example, Facebook\footnote{http://developers.facebook.com/docs/reference/api} and OpenSocial\footnote{http://opensocial.org/} allow 3rd-party application developers and websites to access the social information and activity of millions of users.
However, hidden incentives for users to have many OSN ``friends'' lead to declarations of contacts with little connection in terms of trust, common interests, shared objectives, or other such manifestations of real social relationships~\cite{golder07rhythms}.
Thus, an application that tries to provide targeted functionalities using social information exposed by current OSNs must wade through a lot of noise.

This paper presents Prometheus, a peer-to-peer (P2P) service that collects social information from multiple sources and enables a wide range of new socially-aware applications to mine it via an API that implements social inferences.
This service collects information from actual interactions between users within multiple environments and under multiple identities (e.g., OSNs, email, mobile phones).
Thus, it maintains richer and more nuanced social information than current OSNs or social applications, which can lead to more accurate inferences of trust, interests, and context.
Prometheus represents social information as a directed, weighted and labeled multi-graph distributed on a P2P platform.
Access to such aggregated information is controlled by user-defined policies.

The choice of a P2P architecture was motivated by two factors: user privacy and service availability.
Two alternatives could be considered: a centralized service, and a decentralized service running on mobile phones.
Companies such as Google and Facebook already perform centralized aggregation of user data from independent services.
However, they have no incentives or appropriate business models to allow users full control over their data.
Additionally, some OSNs institute particularly draconian policies concerning the ownership of user-contributed information and content.
For example, users cannot delete their data from the Facebook servers and they cannot easily export their data to a competitive service.
Furthermore, users must trust their OSN provider for complying with privacy policies related to sharing data with 3rd parties.
However, the recent scandal of NSA's unauthorized snooping of millions of users' data from such companies~\cite{theguardian14nsa,washingtonpost13nsa} is yet another proof of Big Brother monitoring.
In contrast, a decentralized service that gives users full control of how their private information is aggregated and accessed is desirable.

An alternative architecture is to store social information in a fully decentralized manner on users' mobile devices, as proposed in~\cite{mokhtar09middlewarepsn,pietilainen09mobiclique,sarigol10tuplespace,toninelli11yarta}. 
In this architecture, a mobile device stores its user's egonet: all the relevant social encounters of its user with others.  
However, research~\cite{benevenuto09characterizingusers,jiang10renren-latentinteractions} found that $30$ to $70\%$ of requests in OSNs, such as for viewing profiles or sending messages, came not from direct social contacts, but from neighbors at least 2 hops away in the social graph.
Such requests could not be satisfied solely from the information stored in egonets.
Thus, more infrastructure would be needed to allow gathering of data from other mobile devices, therefore, including issues related to access, online presence, and limited resources (especially battery).

These aspects led us to opt for an in-between architecture for Prometheus, fully decentralized but capable to aggregate groups of egonets on the same peer and with higher availability.
An overview of the Prometheus service is presented in Section~\ref{sec:overview}, and related work is covered in Section~\ref{sec:related}.
Section~\ref{sec:design} presents a detailed design of the service.
We evaluated Prometheus over large-scale graph simulations and with a large-scale deployment of a prototype on PlanetLab and tested its performance with high-stress workloads from emulated applications.
The results in Section~\ref{sec:exp-eval-merged} show that the response time and system overhead for social inference execution is reduced when user social data are distributed onto peers using a geo-socially-aware approach.
We also implemented CallCensor, a mobile social application that uses Prometheus to decide whether to filter out incoming calls based on the current social context.
As shown in Section~\ref{sec:callcensor}, the response time for this application running on a Google Android phone meets real-time deadlines, demonstrating that Prometheus' overhead is practical for real applications.
In addition, in Section~\ref{sec:resilience} we study the system's resilience to malicious attacks and  demonstrate that its socially-informed design and social graph structure make the system more resilient to graph and inference request attacks.
We conclude in Section~\ref{sec:discussion}.

\section{Overview}\label{sec:overview} 

\begin{figure}[htbp]
\vspace{-2mm}
\begin{center}
	\subfigure[Prometheus in the Social Hourglass Infrastructure]{
		\includegraphics[scale=0.45]{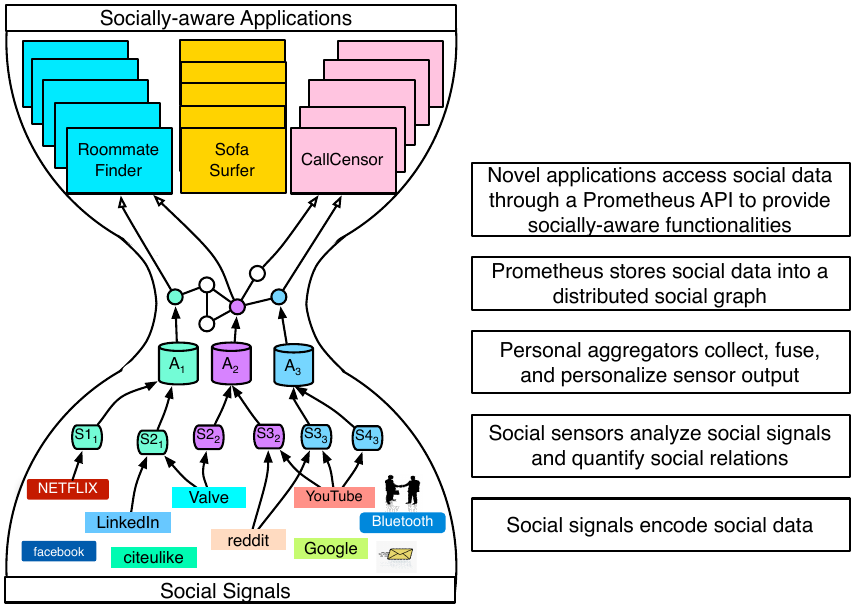}
	}\label{fig:hourglass-prometheus}
	\hspace{-4mm}
	\subfigure[Overview of Prometheus Architecture]{
		\includegraphics[scale=0.53]{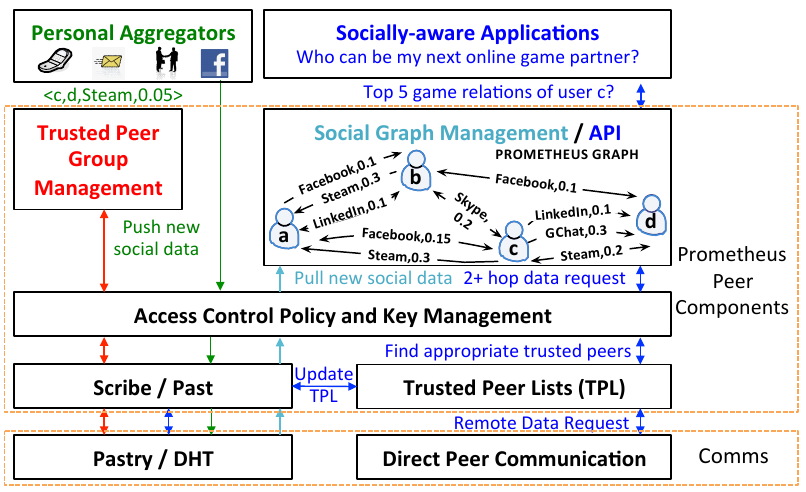}
	}\label{fig:arch}
	\caption{
	(a) Prometheus in the social hourglass infrastructure.
	(b) Prometheus peers are organized using Pastry, a DHT-based overlay, Scribe, a DHT multicast infrastructure, and Past, a DHT storage system.
	Scribe and Pastry are used for social graph and group maintenance.
	Users can trust selected machines, allowed to decrypt and mine the user's social subgraph to fulfill requests.
	Personal aggregators store social input on Past and trusted peers retrieve and represent it as a directed, labeled, weighted multi-graph.
	Applications mine the graph via Prometheus API social inferences, executed between peers through direct P2P communication.}
	\label{fig:hourglass-architecture}
\end{center}
\vspace{-2mm}
\end{figure}

In order to better understand the functionality of the Prometheus service, we present it as part of the \emph{social hourglass infrastructure}~\cite{iamnitchi12hourglass,kourtellis12thesis}.
This infrastructure (illustrated in Figure~\ref{fig:hourglass-architecture}(a)) consists of five main components: (1)~social signals, (2)~social sensors, (3)~personal aggregators, (4)~a social knowledge service (SKS), and (5)~social applications.
Social signals are unprocessed user social information such as interaction logs (e.g., phone call history, emails, etc.) or location and collocation information.
Social sensors run on behalf of a user and parse the user's social signals, analyze them, and send processed social information to the personal aggregator of the user.
The user's personal aggregator combines social information from the user's sensors and produces a personalized output based on user preferences.
This personalized social information is sent for storage and management to the SKS which maintains an augmented social graph that can be mined by applications and services through an API that implements social inferences.
Prometheus fulfills the role of the SKS in the social hourglass infrastructure.

\subsection{Input from Social Sensors \& Aggregators}\label{social-input}

We refer to \emph{social signals} as the information that exposes social interactions between people.
Diverse social signals already exist as byproducts of Internet- or phone-mediated interactions, such as email logs, comments on blogs or OSNs, instant messaging, ratings on user-generated content, phone call logs, or via face-to-face interactions determined from (GPS or Bluetooth-reported) collocation data.
For example, a social signal could reflect the interactions of two users over a soccer video posted on YouTube.
These interactions reflect comments, ``likes'', other video sharing, etc.

\emph{Social sensors} analyze users' social signals.
Sensors are applications running on behalf of a user on various platforms such as their mobile phone, \textsc{pc}, web browser, or trusted 3rd party services.
They transform the domain-specific interactions between their user-owner (\emph{ego}) and others (\emph{alter}) into a message with the format: $ego:<$$alter, label, weight$$>$, where $label$ specifies the interaction domain (e.g., $Facebook$, $Gaming$, etc.) and $weight$ associates a numeric value (e.g., in range $[0,1]$) to the interaction's intensity, as in~\cite{xiang10modeling}.
Many such sensors already exist, although they may not output social ties as they have been implemented in different contexts and for different purposes.
For example, such sensors record and quantify OSNs' user activity~\cite{lewis08tastes}, co-appearance on web pages~\cite{matsuo06polyphonet}, or co-presence recorded as collocation via Bluetooth~\cite{eagle06sensing}.

The \emph{personal aggregator} is a trusted application typically running on a user-owned device (e.g., mobile phone) responsible for fusing and personalizing the information from the user's social signals.
It can perform sophisticated analysis on the input from the sensors, for example, to differentiate between routine encounters with familiar strangers and interactions between friends~\cite{eagle06sensing}.
It also applies user preferences for weighing signals differently: for example, a Google Talk social signal may weigh less than Skype Chat in the aggregated value.
In effect, the aggregator transforms both the labels and weights received from sensors into a quantitative representation of \emph{ego} and \emph{alter}'s social interactions.
Finally, it sends to the SKS a tuple for an aggregated interaction between $ego$ and an $alter$, in the format:
$<$$ego, alter, general\_label, aggregated\_weight$$>$.

\subsection{Prometheus as a Social Knowledge Service}\label{sks}

The SKS provides a mechanism for storing and managing user social data and exposing them to applications and services.
Prometheus, which fills-in the role of SKS in the social hourglass infrastructure, distributes the social information received from the aggregators on multiple user-contributed peers for better service and data availability, while protecting access to data via encryption and user-specified access policies. 
Prometheus design supports the following functionalities.

\emph{Persistent service for mining the social graph} is achieved by employing a p2p architecture on user-contributed resources without the availability constraints of architectures based on mobile phones.
In particular, our objective is to allow applications access to distant nodes in the social graph, such as friends-of-friends queries. 

\emph{Support to an unrestricted set of social applications and services} is provided via two mechanisms: an API that exports social inferences (such as social strength between indirectly-connected users) and aggregated social information represented by social edge labels and weights.
For the latter, Prometheus maintains a weighted, directed, labeled, and multi-edged social graph, where vertices correspond to users and labeled edges correspond to interactions between users, as reported by their aggregators (see Figure~\ref{fig:hourglass-architecture}(b) for an example of such a graph).
Edge labels specify the type, and weights specify the intensity of interaction.
Edge directionality reflects the asymmetry of users' perception of a social relationship~\cite{wellman88analysis} and allows for potentially conflicting reports from the interacting users' aggregators.

\emph{User-controlled data access} is guaranteed via two mechanisms.
First, data owners can specify access control policies to limit access to their information. 
Second, all data are encrypted; however, for mining the social graph information, some peers need to decrypt data in order to process it. 
Thus, each user selects a group of \emph{trusted} peers; only these peers can decrypt the user's social data.

In addition, Prometheus deals with the typical challenges of churn-prone p2p systems by leveraging well-established, previously proposed solutions in distributed hashtable overlays (DHTs).
Figure~\ref{fig:hourglass-architecture}(b) presents an overview of the architecture.
Each peer runs three software components: 1)~for social graph management and application request execution; 2)~for access control policy and key management; and 3)~for trusted peer group management.

\subsection{Output to Applications and Services}\label{social-output}

Prometheus exposes an interface to a rich set of social inference requests computed over the distributed social graph.
For example, an application can request on its user's behalf to receive the top $n$ relations with whom the user interacts over Facebook.
Similarly, it can request the social strength between a user and another user not directly connected in the social graph. 
Prometheus provides the mechanism with which inference requests can access not only a single user's social neighborhood (i.e., directly connected users), but also the social data of users located several hops away in the global social graph.
As mentioned before, all inferences are subject to user-defined access control policies enforced by the trusted peers of the data owner.

\subsection{Prometheus in Use: GamePartnerFinder Application}\label{scenario}

An application scenario is presented below for explaining how Prometheus is used in the context of the social hourglass infrastructure.
Let us assume that Tom is an active online gamer.
However, due to his type of work, he frequently travels around the world and his change of location does not allow him to easily find game partners to play online: he is always on different timezone with respect to his home buddies and also network delays can greatly hinder his gaming experience.
To solve this issue, he developed the $GamePartnerFinder$, an application which utilizes his geo-social information stored on Prometheus to find for him partners to play with while on travel.
The application takes into account information such as: (1)~his friends on Steam Community ($SC$) he frequently plays with, (2)~his friends on Facebook ($FB$) he interacts with, (3)~his frequent phone / Google Chat / Skype contacts with respect to IM and calls, (4)~his professional contacts on LinkedIn, (5)~his current location and friends he is collocated with (at a city / country / continent level).

At installation, the application checks with Tom's aggregator which of the required and optional social sensors he is currently registered with.
These sensors report Tom's activity to Prometheus, subject to a personalized filter stored and applied by his aggregator (as mentioned earlier).
This filter is updated rarely---when there are significant changes in activity patterns or new social sensors are deployed---or can be left to a default setting that weighs all signals equally.
$GamePartnerFinder$ will mine Prometheus' aggregated information on Tom in order to decide which contacts to present as possible game partners for his next session at his current location.

To compile this list, $GamePartnerFinder$ will first query Prometheus for a list of Tom's social contacts that are also geographically close to him.
This allows the application to infer which contacts would be good candidates with respect to network delays and timezone (e.g. same geographical region).
Further, the application will query Prometheus for the type of connection and social distance between each returned contact and Tom.
Contacts classified as ``friends of friends'' (i.e., 2-hop neighbors) could be included in the list, depending on user-specified application-related preferences.
For instance, Tom could specify that 1-hop $SC$ or $FB$ friends can be included, as well as 2-hop $SC$ friends connected over high activity weights, but not 2-hop $FB$ friends who are not already in the previous categories.
LinkedIn contacts could be filtered out.
The multi-hop request allows the application to infer a diverse list of online contacts, close to Tom with respect to network delays, with similar game expertise, and have the social incentives to accept a game invitation from Tom.

Non time-critical applications such as the $GamePartnerFinder$ could wait for some time (e.g., minutes) before producing a good list of game partner candidates, especially if this search involves multi-hop inferences and network delay measurements.
However, more time-critical applications such as a context-aware phone-call filtering application (e.g., the CallCensor~\cite{kourtellis10prometheus,kourtellis10prometheus-nsdi} and Section~\ref{sec:callcensor}) could have a lower tolerance on waiting time.
To facilitate both types of applications, we extended the Prometheus API from past work to support not only the above geo-social requests, but also allow applications to define a timeout parameter which limits the application waiting time.
Thus, if Bob installed $CallCensor$, his incoming personal calls could be automatically silenced during professional meetings, but co-workers' or other professional-related calls could be let through, and this filtering decision should be made within a very short time period (e.g., a few seconds) before the call is dropped or goes to voicemail.

\section{Related Work}\label{sec:related} 

Decentralized management of social data using peer-to-peer architecture has been proposed in other studies such as PeerSoN~\cite{Buchegger09Peerson}, Vis-\`{a}-Vis~\cite{shakimov09vistradeoffs}, Safebook~\cite{cutillo10safebook}, LifeSocial.KOM~\cite{graffi11lifesocialkom} and LotusNet~\cite{aiello10lotusnet}.

PeerSoN~\cite{Buchegger09Peerson} is a two-tier system architecture: the first tier is used to lookup (through a DHT) for metadata of users, current location/IP, notifications, etc.; the second tier allows opportunistic and delay-tolerant networking and is used for the actual contact between peers and users and the exchange of data (files, profiles, chat, etc.).
Users can exchange data either through a DHT storage or directly between their devices.
Vis-\`{a}-Vis~\cite{shakimov09vistradeoffs} introduces the concept of a Virtual Individual Server (VIS) where each user's data are stored on a personal virtual machine.
While similar to the trusted peer concept in Prometheus, VISs are hosted by a centralized cloud computing provider to tackle peer churn, whereas Prometheus provides a truly decentralized, free of cost infrastructure and uses social incentives to reduce churn of user-contributed peers.

Safebook~\cite{cutillo10safebook} is a decentralized OSN with a 3-component architecture: a trusted identification service to provide each peer and user unambiguous identifiers, a P2P substrate for lookup of data, and ``matryoshkas'', concentric rings of peers around each member's peer that provide trusted data storage, profile data retrieval and obscure communication through indirection.
Similarly, Prometheus users employ multiple trusted peers and forwards messages between peers based on social relations and inference requests, while restricting a large scale view of the graph through P2P decentralization. 
Prometheus differs, however, in the exposure of the graph as a first class data object through inference functionality.

LotusNet~\cite{aiello10lotusnet} is a DHT-based OSN that binds user identities with their overlay nodes and published resources for increased security.
Users control access to their data by issuing signed grants to other users applied during retrieval of data.
Services such as event notification, folksonomic content search and reputation management are demonstrated.
In LifeSocial.KOM~\cite{graffi11lifesocialkom}, user information is stored in the form of distributed linked lists in a DHT and is accessible from various plugin-based applications, while enforcing symmetric PKI to ensure user-controlled privacy and access.
User data are isolated from one another, and peers access them individually.
Commercial efforts such as
Enthinnai,\footnote{http://www.enthinnai.com/} FreedomBox,\footnote{http://freedomboxfoundation.org/} and Diaspora\footnote{https://joindiaspora.com/}
implement distributed social networking services allowing users to share content in a decentralized fashion.

Our work differs from the previously noted academic and commercial approaches in the following ways.
First, it not only collects and stores user social information from multiple sources in a P2P network, but also exposes social inference functions to socially-aware applications that mine the social information stored.
Second, it addresses the space between centralized and fully decentralized systems, as it allows multiple users to store social information on common peers they trust.
These peers can locally mine the collection of social data entrusted to them by the groups of (possibly socially-connected) users, thus improving social data access and quality of inferences to applications.
Third, Prometheus enables users to select trusted peers independent from the users' (mobile) devices to decrypt and mine their social data at any time, thus increasing service availability.

\section{Design}\label{sec:design}

Prometheus manages a labeled, weighted social multigraph distributed on a collection of user-contributed peers. 
Unlike other p2p solutions, Prometheus does not require each user to contribute a peer. 
Instead, it leverages social incentives to allow a user's data be managed by the peer contributed by a trusted social relation.
This approach embeds social awareness in the service design and has three immediate benefits. 
First, it improves scalability since the system size can be much smaller than the number of its users.
Second, socially-incentivized users keep their computers online, thus reducing churn~\cite{li06f2f,tran08friendstore} and consequently increasing service availability.
And finally, because socially-related users are likely to select the same trusted peers for their data management, the resulting data collocation decreases remote access for social graph traversals.
A less intuitive benefit from the social-based mapping of the social graph onto the p2p network includes a reduced vulnerability to malicious attacks in comparison to random-based mappings such as pure DHT-based systems, as demonstrated in Section~\ref{sec:resilience}.

Prometheus uses a public-key infrastructure (PKI) to ensure both message confidentiality and user / peer authentication.
Using their personal trusted device (e.g., mobile phone or laptop), users create public and private keys for both themselves ($P$, $p$) and their trusted peer group ($G$, $g$) decrypting their social data.
The group public key ($G$) can be placed on the DHT for other peers to access for decryption of incoming messages.
Peers trusted by a user must have his group keys as well as his personal public key ($P$).
All messages contain nonces added by the senders.
They are reattached in the replies by the receivers to acknowledge correct delivery and execution, and in combination with timestamps, they dissallow replay attacks.
After these appendices, the messages are digitally signed with the sender's private key to preserve integrity and confirm authenticity, and then encrypted with the receiver's public key to protect the message from unauthorized receivers.
The Prometheus social data consist of labeled and weighted social edges as reported by personal aggregators.
They are signed, encrypted and stored in Past~\cite{rowstron01past}, a DHT-based storage built on top of the Pastry overlay~\cite{rowstron01pastry}.
Only trusted peers are capable of decrypting their user's data for graph processing.

\subsection{User Registration}\label{sec:registration}
A user registers with Prometheus by creating a uniform random $UID$ from the circular 128-bit ID space of the DHT (typically the hash of their public key).
At registration time, the user specifies her peer(s) contributed to the network (if any).
The user then creates a mapping between her $UID$ and the list of these peers' IP addresses, signs it with her private key and stores this mapping in the DHT network as the key-value pair $UID=\{IP_1, ...,IP_n\}$.
When one of these peers returns from an offline state, it updates the mapping with its current IP address.
The user also compiles a list of other Prometheus users with whom she shares strong out-of-band trust relations (and has exchanged public keys) and retrieves from Past their machine mappings.
She confirms the mappings' authenticity and from the returned list of peers, she selects an initial set as her trusted peer group.
The larger this set, the higher the service availability; at the same time, the consistency and overall performance may decrease.
The trusted peer group may change over time, as shown next.

\subsection{Trusted Peer Group Management}\label{group_management}

The main issues concerning the trusted peer group management are group membership and search for trusted peers.
For managing the trusted peer group we leverage Scribe~\cite{castro02Scribe}, a group communication solution designed for DHTs.

In general, communication between trusted peers for group management happens over the DHT/Scribe's multicast protocol by appending the message to be sent with the groupid, signing with the group private key and encrypting with the group public key: $\{\{msg,nonce,groupid,timestamp\}g\}G$.
The receiving trusted peers decrypt the message with the group private key, verify the signature of the sender peer as trusted using the group public key and read/execute the message.

The group management happens as follows.
Let us consider that Alice uses her personal device as a proxy to add peers to her trusted peer group, initialized as a Scribe group with the handle ``Trusted\_Peer\_Group\_UID'', where UID is Alice's user id.
For this, she initiates a secure three-step handshake procedure to establish a two-way trust relationship between her and Bob, a peer owner.
The following steps take place in the direct peer communication layer (Figure~\ref{fig:hourglass-architecture}(b)):
(1)~Through out-of-band methods, she has invited Bob and he has accepted to participate in her trusted peer group by sharing his public key (as mentioned earlier). 
(2)~Alice sends Bob the keys of her trusted peer group (enhanced with a nonce, timestamp, Bob's userid, and then signed with her private key and encrypted with Bob's public key).
(3)~By decrypting the message with his private key and verifying its authenticity with Alice's public key, Bob installs on his peer the received group keys, enabling his peer to join Alice's group by sending a digitally signed and encrypted \emph{subscribe} request to Alice's Scribe trusted group:$\{\{subscribe,nonce,groupid,timestamp\}g\}G$.
Receiving these keys allows the newly trusted peer to access the user's social data as explained later.
Note that a malicious peer owner cannot force anyone to give him their data if they don't trust him first.

For removing a trusted peer, Alice will have to create new group keys, re-encrypt her signed data with the new group public key, and store them in the DHT.
The removal of the peer from a user's trusted group is \textit{multicasted} to all group peers and the peer is unsubscribed from the group.
The new group keys are distributed to the still trusted peers as before.
The distribution of new keys and re-encryption of the social data disallows the removed peer from decrypting updated versions of the user's data.
If a peer owner decides to remove her peer from another user's trusted group (e.g, due to overload), the removal request is multicasted to the group, which in turn executes the above procedure and alerts the data owner of the change.

A requesting peer or application can find a user's trusted peers by submitting a \textit{multicast} TPL request with handle \textit{Trusted\_Peer\_Group\_UID}.
With the multicast, all online group trusted peers respond with their IP and signed membership, which is verified for authenticity with the user's group public key.
The requesting peer creates a trusted peer list (TPL) of IPs based on peer responses.
The multicast allows the requesting peer to have peer alternatives in case of churn or erroneous communication with the first responding peer.
The search for trusted peers follows the Scribe multicast tree which utilizes the Pastry DHT routing and its network proximity metric to forward messages to near-by peers, thus reducing overall delays.
In effect, the responds from trusted peers are sorted in the TPL by overall latency.
The peer, upon creating the TPL for a user's group, can communicate directly with the individual trusted peers, preferably the one responding fastest.
Prometheus caches the TPL after the first access and updates it when the trusted peers are unresponsive, changed their IP, or became untrusted.
Users can also apply their own TPL update policies based on their usage patterns.
A user's social data are unavailable if all her trusted peers leave the network; no service requests involving this user can be answered until a trusted peer rejoins the network.
However, we ensure durability of users' generated data (i.e., input from social aggregators) via encrypted storage and replication in the DHT by Past.

\subsection{Distributed Social Graph}\label{dist-social-graph}

\begin{figure}[t]
\centering
	\includegraphics[scale=0.5]{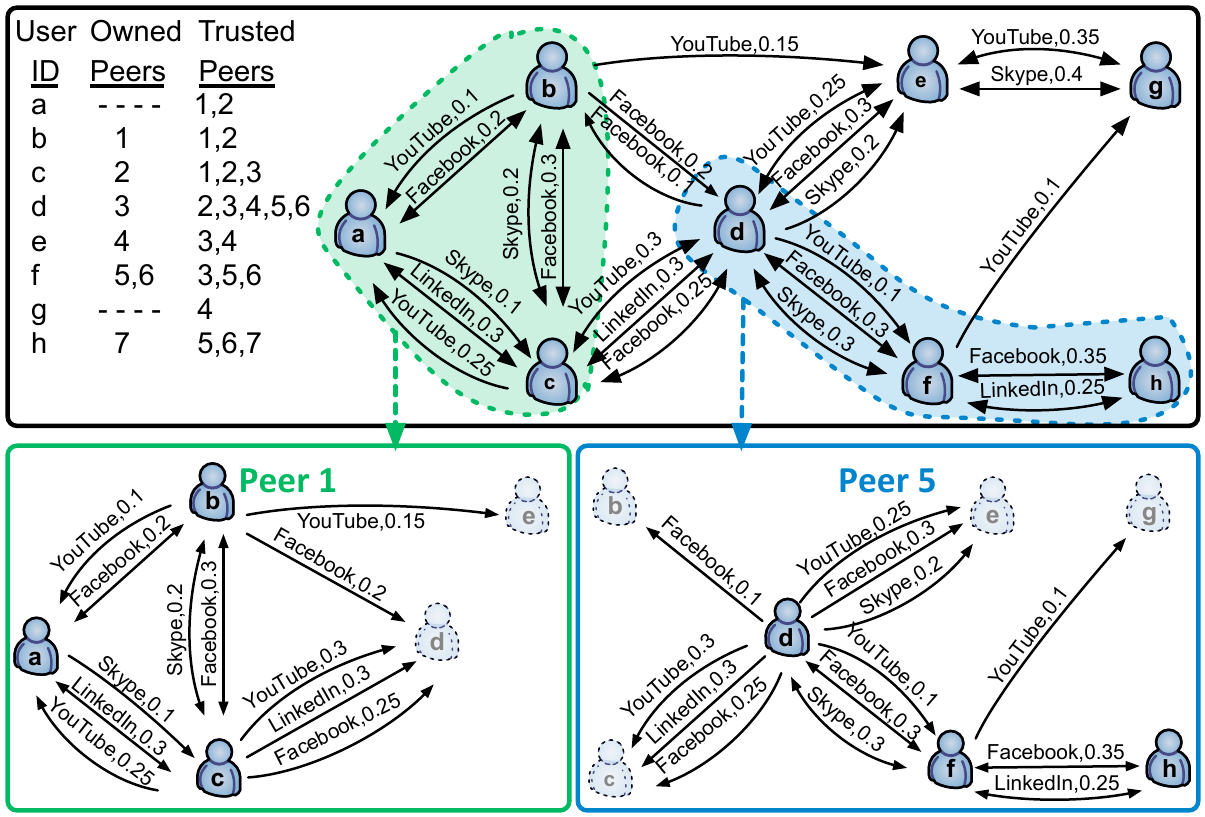}
	\caption{An example of a social graph for eight users (\textit{a-h}) distributed on seven peers.
	The left figure shows the mapping between users, peer owners, and trusted peers (upper left corner) and how users are connected with each other over social edges, each marked with its label and weight.
	The right figures illustrate the subgraphs maintained by peers \textit{1} and \textit{5}.
	Users in dark color (e.g., $a, b, c$ on peer \textit{1}) trust the peer to manage their social data.
	Users in light-shaded color (e.g., $e, d$ on peer \textit{1}) do not trust the peer but are socially connected with users who do.
	}\label{fig:Example}
\end{figure}

Prometheus represents the social graph as a directed, labeled, and weighted multi-edged graph~\cite{paanders10sns}, maintained and used in a decentralized fashion as presented in Figure~\ref{fig:Example}.

Multiple edges can connect two users, and each edge is labeled with a type of social interaction and assigned a weight (a real number in the range $[0,1]$) representing the interaction's intensity.
The labels for interactions and their associated weights are assigned by the personal aggregator of each user.
From an application point of view, distinguishing between different types of interactions allows for better functionality.
For extensibility, we designed Prometheus to be oblivious to the number and types of social activities reported by the aggregators.
Furthermore, we chose to represent the graph as directed and weighted because social ties are asymmetrically reciprocal~\cite{wellman88analysis}.
This representation also limits illegitimate graph uses (e.g., for spamming).
A user's latest known location and an associated timestamp are maintained as an attribute of the user's vertex in the graph.

Social data (sd) of a user are appended with a nonce, his group's id and a timestamp, then signed with the user's private key and then encrypted with the user's group public key: $\{\{sd, nonce, groupid, timestamp\}p\}G$.
This encrypted item is stored in Past in the append-only file $Social\_Data\_UID$.
Only the user's personal aggregator can send updates to create new edges, remove edges, or modify edge weights.
These updates are appended in this file as records with a sequence number.
Trusted peers periodically check the file for new records and retrieve all such records: this is easily done based on sequence number comparison starting from the end of the file.
They decrypt new records with the user's group private key (g) and verify the data's authenticity with the user's personal public key (P).
By checking the nonce, they confirm they haven't received this item before and incorporate the changes with the rest of the local graph.
As the social data file is append-only, the trusted peers can access it at any time: in the worst case, they will miss the latest update.
Therefore, for short periods of time, the trusted peers may have inconsistent data, but this is not a major problem as social graphs do not change often~\cite{golbeck07osnchange}.

Edges may ``decay'' over time if few (or no) activity updates are received~\cite{roberts10decay}.
This aging process should be activity specific, but it should also reflect the user's social habits and interests: users who are less socially active and users with a great number of friends should have their relationships age slower.
Currently, the system applies a simple aging function to reduce an edge's weight by $10\%$ for every week the two users do not interact over the particular label (thus, the connection never completely disappears and the aging happens slowly).
A user's aggregator may specify a different decrement value of the weight and the time period for aging (these values are also stored in the $Social\_Data\_UID$ file for each user).

\subsection{Social Inference API}\label{inference-api}

Prometheus exposes to applications an API of social inference functions that are executed in a decentralized fashion; more complex inferences can be built on top of this set.

\emph{Relation\_Test(ego, alter, $\alpha$, $\chi$)}
is a boolean function that checks whether $ego$ is directly connected to $alter$ by an edge with label $\alpha$ and with a minimum weight $\chi$.
The CallCensor application~\cite{kourtellis10prometheus,kourtellis10prometheus-nsdi} can use this function to determine if an incoming call is from a coworker with a strong social tie, and therefore, should be let through even on weekends.

\emph{Top\_Relations(ego, $\alpha$, n)}
returns the top $n$ users directly connected to $ego$ by an edge with label $\alpha$, and ordered by decreasing weights.
An application can use this function, for example, to invite users highly connected with $ego$ to share content related to activity $\alpha$.   

Our previous study~\cite{kourtellis10prometheus} revealed volatility of the peer-to-peer communication and long response delays during multi-hop inference execution.
Thus, we redesigned the following API functions to offer better quality of service to applications, by allowing them to define a timeout parameter $T$, which limits the application waiting time.

\emph{Neighborhood(ego, $\alpha$, $\chi$, radius, T)}
returns the set of users in $ego$'s social neighborhood who are connected through social ties of label $\alpha$ and minimum weight $\chi$ within a number of social hops equal to \textit{radius}.
The \textit{radius} parameter allows for a multi-hop search in the social graph (e.g., setting \textit{radius} to 2 will find $ego$'s friends of friends).
The GamePartnerFinder application can use this function to determine if a user is in \textit{ego}'s ``gaming'' graph neighborhood even if not directly connected.

\emph{Proximity(ego, $\alpha$, $\chi$, radius, distance, timestamp, T)}
is an extension of the neighborhood function which filters its results based on physical distance to $ego$.
After $ego$'s location information is collected, and $neighborhood$ returns a set of users, $proximity$ returns the set of users within $distance$ from $ego$ and their location information is at most as old as $timestamp$.
Users who do not share their location or have location information older than the $timestamp$ are not returned.
A mobile phone application like CallCensor might use this function to infer the collocated coworkers within $distance$ from $ego$.

\emph{Social\_Strength(ego, alter, T)}
returns a real number in the range $[0,1]$ that quantifies the social strength between $ego$ and $alter$ from $ego$'s perspective.
Past studies in sociology~\cite{friedkin83horizons} observed that two individuals have meaningful social relationships when connected within two social hops (i.e., ``horizon of observability''), and their relationship strength greatly depends on the number of different direct or indirect paths connecting them.
Therefore, the two users in this function can be directly or indirectly connected and through multiple parallel paths (e.g., paths $c$$\rightarrow$$b$$\rightarrow$$e$ and $c$$\rightarrow$$d$$\rightarrow$$e$ between $c$ and $e$, Figure~\ref{fig:Example}).
The returned value is normalized, as explained below, to $ego$'s social ties, to ensure that the social strength is less sensitive to a particular user activity.

Assume $\Lambda_{i,j}$ is the set of labels of edges for two directly connected users $i$ and $j$, $w(i,j,\lambda)$ is the weight of an edge between $i$ and $j$ over label $\lambda$, and $\Theta_i$ the set of directly connected neighbors to $i$.
Then, $nw(i,j)$ (eq.1) is the overall normalized weight between $i$ and $j$.
Also, assume users $i$ and $m$ are indirectly connected over different $2$-hop paths.
Then, $S(i,m)$ (eq.2) is the value returned for social strength between $i$ and $m$, over a multi-level 2-hop path.
This function could be used, for example, to assess social incentives for resource or data sharing.

\vspace{-2mm}
{\footnotesize
\begin{equation*}
nw(i,j)=
\frac
{
\displaystyle\sum_{\forall \lambda \in \Lambda_{i,j}}
w(i,j,\lambda)
}
{
\displaystyle\max_{\forall j \in \Theta_i}
\Bigl(
\displaystyle\sum_{\forall \lambda \in \Lambda_{i,j}}
w(i,j,\lambda)
\Bigr)
}
\;(1),
\qquad
S(i,m)=
1-
\!\!\!\!\!\!
\prod_{
\substack{
\forall j \in \Theta_i \cap \Theta_m
}
}
\!\!\!\!\!\!
\left(
1-
\frac
{
\min
\{
nw(i,j), nw(j, m)
\}
}
{2}
\right)
(2)
\end{equation*}
}
\vspace{-2mm}

\textbf{Inference Function Execution:}
Assume user $x$ wants to submit (through an application) a social inference request to one of her trusted peers.
Using the process explained in Section~\ref{group_management}, the application creates $x$'s TPL and forwards the request to $x$'s fastest responding and thus network-closest (shortest latency) trusted peer.
This is done through the direct communication layer in a secure fashion by appending on the request a nonce, $x$'s userid and a timestamp, then signing the message with $x$'s private key and encrypting it with $x$'s group public key: $\{\{request,userid,nonce,timestamp\}p\}G$.
The receiving peer decrypts the message with $x$'s group private key (g), verifies the submitter's identity via her public key and if the query has not been seen before (by checking the nonce) and the peer is indeed trusted by $x$, the peer enforces $x$'s access control policies (described next).
In all other cases, the request is dropped.
Then the peer fulfills the request by (1)~traversing the local social subgraph for the requested information, (2)~signing and encrypting the result using the requesting user's public key and the group's private key, respectively, and (3)~returning it to the user (application).

For functions that can traverse the graph for $n$ hops (i.e., \emph{Neighborhood}, \emph{Proximity} and \emph{Social\_Strength}), the peer submits signed and encrypted secondary requests for information about other users to their trusted peers, as follows.
A secondary request includes the $UID$ of the original submitter in order to verify her access rights, then it is signed with her group private key and encrypted with the receiving user group's public key.
A time period of $T*(n-1)$ seconds is given to the secondary peers to respond with their results (peers at each $hop$ use independent clocks for $T$ seconds).
Each receiving secondary peer decrypts the request with the group private key, authenticates the signature with the requesting group's public key and checks the access control policies for the requesting user.
If the request is granted, the result is returned to the requesting peer.
If the request still needs more information, that peer (e.g., at hop $k$) repeats the same process and submits a secondary request with an adjusted $T*(n-k-1)$ timeout.
Finally, the original requesting peer recursively collects all the replies and submits the final result to the application.

\subsection{Access Control Policies}\label{acps} 

Users can specify access control policies (ACPs) upon registration and update them any time thereafter.
These policies, stored on each of the user's trusted peers, are applied each time an inference request is submitted to one of these peers to access social information for the particular user.
For availability, the policies are signed, encrypted and stored in the DHT,
allowing only a rejoining trusted peer to recover updated policies (similarly with the social data).

ACPs are comprised of two parts: the social data object(s) to be accessed and the specification(s) to be met before access is granted for the particular data object(s).
They are defined as entries of the $ACP\_UID$ file in the following format:
\emph{$<$Social Data Object(s)$>$ :: $<$ACP Specification(s)$>$}.
Table~\ref{tab:acp-definitions-example} presents a list of such access control policy definitions.
By design, ACPs are whitelists.
Any of the specifications can be used to allow access to any of the data objects.
Each of the data objects, as well as each of the specifications, can be combined with logical operators (e.g., \emph{AND}, \emph{OR}, \emph{NOT}, etc.), to create more complex access policies.
To verify the access rights, Prometheus may call its inference functions, when applicable.
For example, to detect whether the request's originating user is within $n$-hops, she is checked against the result of an $n$-hop {\it neighborhood} inference.
ACPs also allow for blacklisted users (and their peers) for convenience.
Given that it may take some time to discover malicious users or peers and update blacklists across the system, some attacks could take place.
As future work, we plan to investigate the effective discovery and blacklisting of malicious peers, as well as the provision of strong consistency and conflict resolution for policies.

Table~\ref{tab:acp-definitions-example} also shows an example of the set of access control policies for $Bob$.
He allows \emph{work}-related social information to be given to requests coming from users within $2$ hops who are connected with him over \emph{LinkedIn} label, or when these requests come from the \emph{CallCensor} application.
Also, he allows his parents and his brother to access his exact location at any time.
If a neighborhood inference request for the \emph{Facebook} label is submitted to $Bob$'s trusted peer, Prometheus checks his ACP in the order \textit{Blacklist$\rightarrow$labels$\rightarrow$weights}.
$Alice$ and $Gary$ are excluded from all types of inferences and cannot receive any information about $Bob$.

To protect their data from illegitimate access, users can set strict ACPs, for example, serving only requests originating from trusted $1$-hop friends.
Such ACPs will disallow $2$-hop contacts from accessing any data, but will also restrict the usability of multi-hop inferences such as $social\_strength$ or $neighborhood$.
Overall, we anticipate a tradeoff between how ACPs are defined to protect social data, and the usability of the platform through social inferences on these data.
However, as demonstrated in the ACPs definition, users can fine-tune them to allow different access rights to particular sets of users, enabling inferences, and in extend applications, to maintain their functionality by accessing specific portions of user data.

\begin{table}[htbp]
\vspace{-4mm}
\centering
\caption{Access Control Policy Definitions and Examples.\label{tab:acp-definitions-example}}{
\footnotesize
\tabcolsep=0.12cm
\begin{tabular}{l l l}
\toprule
\multicolumn{1}{c}{Social Data Objects}&	\multicolumn{1}{c}{ACP Specifications}	&	\multicolumn{1}{c}{ACP Examples}															\\
\midrule
Social edge label \hfill	$\alpha$	&	Social distance \hfill 		$\rho$		&																						\\
Social edge weight \hfill	$\chi$	&	Social edge label \hfill 	$\gamma$	&	{\scriptsize $< \chi=0.3 > :: < \rho=1 \text{ AND } S=SofaSurfer >\text{ OR } C=Charles) >$}				\\
User location \hfill		$\Delta$	&	Social edge weight \hfill 	$y$			&	{\scriptsize $< \Delta > :: < B=mom \text{ OR } B=dad \text{ OR } B=brother >$}							\\
							&	Originator user	\hfill 		$B$			&	{\scriptsize $< \alpha=Skype > :: < \rho=2 \text{ AND } \gamma=Skype \text{ AND } y=0.2>$}				\\
							&	Originator peer	\hfill 		$P$ 			&	{\scriptsize $< \alpha=Facebook \text{ AND } \chi=0.2 > :: < \rho=1 \text{ AND } \gamma=Facebook>$}		\\
							&	Intermediate user \hfill 	$C$			&	{\scriptsize $< \alpha=LinkedIn > :: < (\rho=2 \text{ AND } \gamma=LinkedIn) $}						 	\\
							&	Intermediate peer \hfill 	$M$			&	{\scriptsize $\text{ \qquad \qquad \qquad \qquad \qquad \qquad OR } S=CallCensor >$}					\\
							&	Application \hfill 		$S$			&	{\scriptsize $< blacklist > :: < B=Alice \text{  OR  } B=Gary >$}										\\
							&	Originator's location	\hfill 	$L$			&																						\\
\bottomrule
\end{tabular}}
\vspace{-2mm}
\end{table}

\section{Performance Evaluation}\label{sec:exp-eval-merged}

To evaluate our system design, we used both simulations (Section~\ref{sec:sim-exps}) and experiments with a real implementation deployed on PlanetLab (Section~\ref{sec:pl-exps}).

\subsection{Simulation Experiments}\label{sec:sim-exps}

We measured system performance, scalability and benefits from caching and geographic distribution of data on peers with three sets of experiments.
In these tests, we simulated the inference functionalities of Prometheus while varying the size of stored social graphs, the type of data mapping on peers, the distribution of network delays, and data caching.

\subsubsection{Experimental Setup}\label{sec:sim-exp-setup}

The two metrics used in this evaluation were {\it expected response time} to quantify the application-perceived performance, and {\it number of P2P network messages} to quantify the service overhead.
We used synthetic bidirectional social graphs created with a generator~\cite{sala10graphmodels} which consistently produces graphs with properties such as degree distribution and clustering coefficient similar to real social graphs, but independent of the particularities of specific real graphs (Table~\ref{tab:synth-nets}).
Edge weights were equally set to $0.1$.
We assumed open ACPs for all users, allowing access to all data from all users, thus stressing the system at maximum load.
We chose a random workload of $10\%$ of graph users requesting data within $1$--$3$ hops, to investigate the system performance equally across all types of users (periphery and popular users alike).

We examined a random and a social-based mapping of user data on peers.
The social mapping corresponds to a realistic scenario where a group of socially-connected users share the resources of a peer contributed by a group member.
For this mapping, the synthetic graphs were split into communities using the Louvain community detection method~\cite{louvain08fast}, then each community was mapped on a peer and we applied no data replication.
Depending on the size of the graph used, the number of peers in the system varied between $1.5$ and $166$ thousand peers, with $6$--$8$ users/peer, on average.
In the random mapping, user data are randomly grouped into sets of the same sizes as in the social mapping, and mapped on peers.

We experimented with three network delay distributions for peer messages, simulating different geographical distributions of data.
Using a $PL$-delay trace\footnote{http://ridge.cs.umn.edu/pltraces.html} between $242$ $PL$ servers located in $21$ countries, we extracted distribution delays for random peers, for peers within the same country, and for peers within the same continent.
Thus, at the $m$-th network hop of an $n$-hop request, peers sent requests to other peers with a delay randomly selected from the delay distribution tested.
The maximum delay for the $m$-th hop is used as its dominating network delay.
The expected response time of a request is the sum of these maximum delays across all hops.

We evaluated the benefits of data caching by storing the results seen from each peer and skipping unnecessary secondary requests the local cache could fulfill.
Every $t$ seconds the cache expired, depending on how frequently edges changed as observed on each peer from incoming social inputs.
For simplicity, we assumed caches were always valid, i.e., an inter-arrival time of requests smaller than the cache expiration period.
We validated our simulator using k-regular synthetic networks\footnote{The SNAP library was used (http://snap.stanford.edu/snap/index.html) with rewiring probability $p=0.1$.} with similar properties as the synthetic social networks (same degree $k$ and number of vertices) and confirmed that the results returned (i.e., messages sent/received) match the analytical calculations (in the order of $O(2k^{(m-1)})$ for the $m$-th hop).
We executed each setup 10 times and report statistically significant average values.

\begin{table}[tbh]
\vspace{-4mm}
\centering
\caption{Description of synthetic graphs used. AD: average degree, CC: clustering coefficient, ED: effective diameter, C: Number of communities, ACS: Average community size.}
\vspace{-2mm}

\begin{tabular}{rrrrrrrr}
\toprule
Dataset	&	Nodes		&	Edges			&	AD		&	CC		&	ED	&	C			&	ACS	\\
\midrule
1k		&	\num{1000}	&	\num{5895}		&	11.8		&	0.263	&	5.47	&	132			&	7.58	\\
10k		&	\num{10000}	&	\num{58539}		&	11.7		&	0.219	&	6.56	&	\num{1562}	&	6.40	\\
100k		&	\num{100000}	&	\num{587970}		&	11.8		&	0.207	&	7.07	&	\num{15692}	&	6.37	\\
1000k	&	\num{1000000}	&	\num{5896878}		&	11.8		&	0.204	&	7.76	&	\num{166314}	&	6.01 \\
\bottomrule
\end{tabular}
\label{tab:synth-nets}
\vspace{-4mm}
\end{table}

\subsubsection{Scalability and Caching}

\begin{figure*}[tbph]
\vspace{-2mm}
\centering
	\includegraphics[scale=0.55]{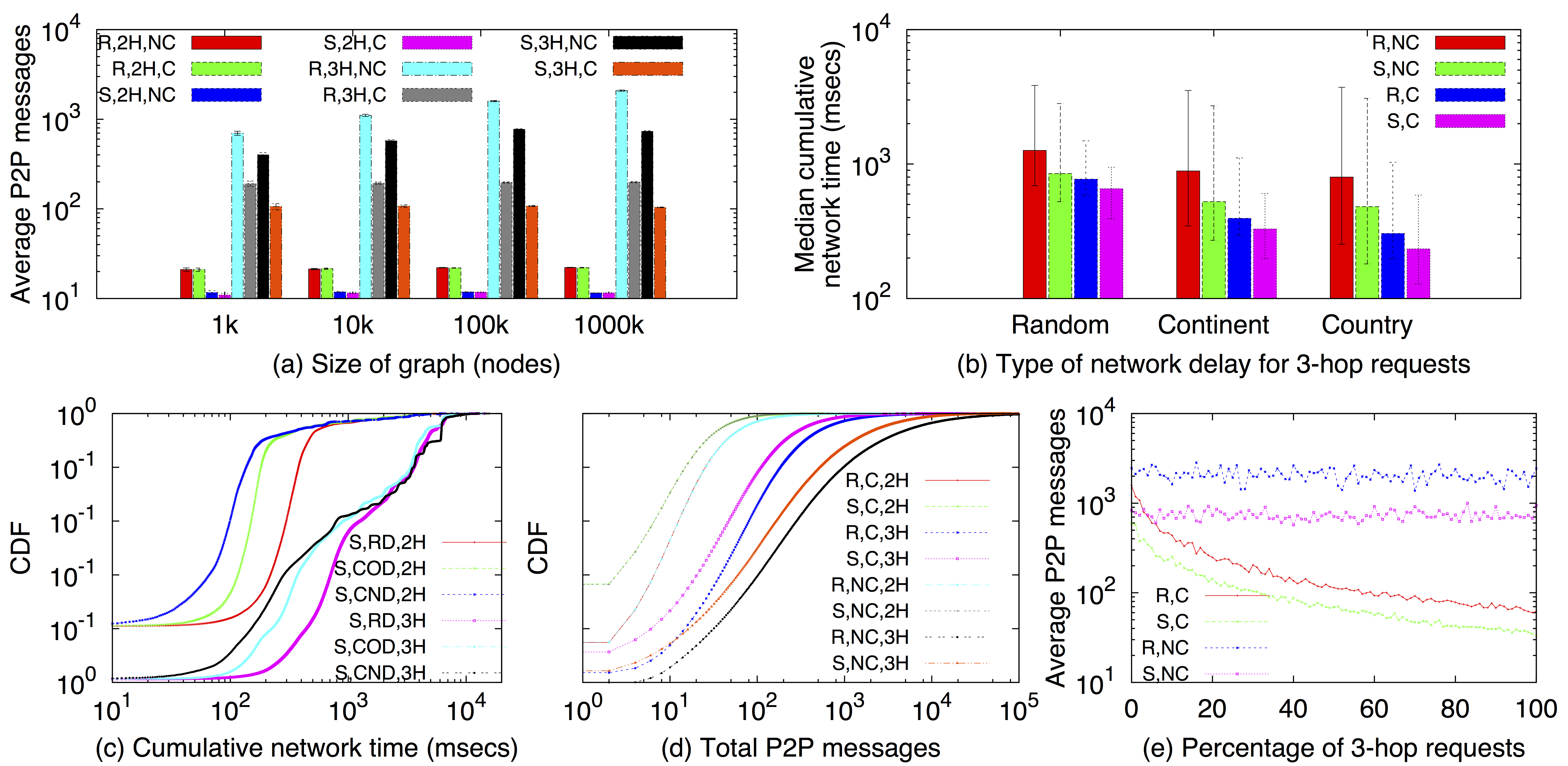}
	\caption{ Simulation results over various settings: 
	Social (S) or Random (R) mappings; 2 (2H) or 3-hop (3H) requests; caching (C) or no-caching (NC); random, continent or country delays (RD, COD, CND).
	(a) Average P2P messages vs. graph sizes (y-bars show $95\%$ confidence intervals).
	(b) Median network times vs. geo-distribution of peers (1M users) (y-bars show $1^{st}$ and $3^{rd}$ quartiles).
	(c) CDF of expected network time with no caching (1M users).
	(d) CDF of total P2P messages (1M users).
	(e) Average P2P messages vs. portion of 3-hop requests served (1M users).}
\vspace{-2mm}
\label{fig:simulation-exps}
\end{figure*}

Figure~\ref{fig:simulation-exps}(a) shows the average number of messages sent between peers over different size graphs, with a detailed CDF for the graph of 1 million users in Fig.~\ref{fig:simulation-exps}(d) (similar distributions were observed for the other graph sizes and are omitted for brevity).
As expected, increasing the graph size forces the system to exchange more P2P messages.
The increase is small for 2-hop requests ($5$-$10\%$) and larger in 3-hop requests ($23$-$40\%$) in random mapping, and ($25$-$30\%$) in social mapping.
The social mapping needs fewer messages: about $50\%$ of random in 2-hop (across all graph sizes) and $40$-$65\%$ of random in 3-hop requests.
When caching is enabled, the 2-hop requests do not benefit since secondary requests sent to other peers start from users connecting different communities, and we never repeat requests from the same users.
Instead, 3-hop requests significantly benefit: when the graph size increases, the increase in messages is reduced to $1$-$2\%$ (from $20$-$40\%$) in both mappings.
However, the social mapping still produces about $44$-$48\%$ fewer messages than random, with the difference increasing with graph size.
Figure~\ref{fig:simulation-exps}(e) shows the effect of caching in subsequent requests: with $50\%$ of requests executed (i.e., $50k$ requests in the $1M$-user graph), the number of P2P messages drops by an order of magnitude.

\subsubsection{Geo-distribution of peers}
Figure~\ref{fig:simulation-exps}(b) shows the median end-to-end response time needed for a 3-hop request to be fulfilled, given different geographical distributions of peers (random, in-continent or in-country).
Also, Figure~\ref{fig:simulation-exps}(c) shows the CDF of the time for 2 or 3-hop requests in the social mapping for the 1M-user graph.
Considering median cases in the random delays (Fig~\ref{fig:simulation-exps}(b)), requests in the social mapping are fulfilled $30$-$40\%$ faster that in random for 3-hops (and $16$-$20\%$ faster for 2-hops, not shown here).
If socially-mapped peers send secondary requests to other peers within the same continent(country), the network times are $50\%$ ($70\%$) smaller than random delays for 2-hops, and about $38\%$ ($45\%$) for 3-hops (Fig~\ref{fig:simulation-exps}(c)).
When caching is enabled (Fig~\ref{fig:simulation-exps}(b)), the social mapping with continent (country) delays reduces response times to $50\%$ ($67\%$) of the random delays in the 3 hops.

\subsection{PlanetLab Experiments}\label{sec:pl-exps}

We implemented Prometheus on top of the FreePastry\footnote{http://www.freepastry.org/} Java implementation of Pastry DHT which also provides API support for Scribe and Past.
We performed various optimizations and fine tuning from our earlier work~\cite{kourtellis10prometheus} to better handle the communication volatility and peer churn typically observed in a distributed infrastructure.
We also redesigned the API to allow applications to submit a serializable class-based request by defining within the class object not only inference-specific parameters (such as label and weight) but also the new timeout parameter, which declares the application waiting time per hop when requesting multi-hop inferences.
Furthermore, applications and peers can define parameters such as request id, timestamp and error flags.
Overall, the new API design enhanced extensibility for future inference parameters, portability across different platforms, and improved the peer-to-peer and peer-to-application communication during inference execution.
We measured Prometheus' performance on PlanetLab ($PL$), a widely distributed network, and assessed the effect of socially-aware trusted peer selection on the system's overall performance.

\subsubsection{Experimental Setup}

In these experiments we deployed the prototype on $100$ selected $PL$ peers worldwide ($35$ countries).
We set the application timeout to $15$ seconds/hop in the first set of experiments and tested lower timeouts in the second set.
This timeout is needed to deal with communication delays due to long RTTs ($125$ms on average), busy network interfaces (the $PL$ peers are shared resources), uncontrolled peer churn, and forwarding of requests.
Prometheus uses this parameter to decide when to aggregate the intermediate results received so far and send them to the requesting peer.
We note that the overall response time measured was dominated by the $PL$ network delays (order of seconds) in comparison to the overhead for different local activities such as request parsing, decryption and signature verification and subgraph traversal on each peer (order of milliseconds); thus we focus 

We did not control the churn; peers were subjected to an average $PL$ churn of $\sim$$5\%$.
To remove any bias from the data placement on faster or slower responding or failing peers, and the submission of requests from or to such peers, we distributed sets of data on randomly selected peers (thus enforcing random delays between peers) and each peer submitted application workload on behalf of all users.
Edge weights were initially set to $0.1$ and dynamically updated over time based on an empirically driven model explained in Section~\ref{emulators_setup}.
In addition to the two metrics reported in simulations, we also measured the {\it percentage of completion} to quantify the tradeoff between request response time and level of request completion.

The overloaded testbed led to increased communication delays while creating trusted peer lists.
For a request for user $x$ to be sent, an application has to first create $x$'s $TPL$---an operation which involves several time-consuming DHT lookups, resulting in multiple peer traversals---and then forward the request to the first responding trusted peer.
The distribution of the overhead associated with the cold start of creating a user's $TPL$ over $PL$ had a $50^{th}$, $90^{th}$ and $99^{th}$ percentile of $1.05$, $2.15$ and $8.78$ seconds, respectively.
Thus, for the majority of users this process can be fast, but for some users it can take as much as $8$--$10$ seconds (similar delays of $6$--$10$ seconds were reported for the Vis-a-Vis' group activities~\cite{shakimov09vistradeoffs} running on $PL$).
To mitigate this problem, Prometheus caches the $TPL$ as explained in Section~\ref{group_management}.
As we are interested not in testing the DHT but Prometheus' performance in inference execution, our results show the performance using this caching mechanism and the direct peer communication layer (as discussed in Figure~\ref{fig:hourglass-architecture}(b)).

\subsubsection{Graph Decentralization}\label{sec:mappings}
We used the $1k$-user graph to reflect a realistic scenario of distributing a social graph on $100$ friend-contributed peers: each contributing user is trusted by a low number of relations.
We vary this number from $10$ to $50$ (allowing replication), although one peer can handle requests for $1000$ users with no performance penalty~\cite{anderson10thesis}.
The scale of the graph and consequently of the P2P network tested is sufficient for evaluation of meaningful social inferences typically traversing a social neighborhood of 2-3 hops.
1000 users mapped on 100 peers provided enough variation for statistically significant measurements on a real deployment.

We distributed the graph on $PL$ peers using a random (as before) and a social mapping.
The average number of users/peer received values $N$=$10$, $30$ and $50$ users/peer.
The number of PL peers was kept constant to $100$, forcing the user groups to overlap for $N$$>$$10$.
This resembles the realistic scenario of overlapping social circles with some users participating in more than one circle (peer), and thus having multiple trusted peers.
In effect, user data were replicated on $K$=$N$/$10$ peers on average, hence $K$=$1$, $3$ and $5$ trusted peers/user.

To produce overlapping communities in the social mapping in a controlled fashion leading to the specific replication factors, we modified the community detection algorithm from~\cite{girvan02community} to control the number of communities and their average size.
The algorithm takes as input the graph, the number of communities to be identified and the minimum acceptable community size.
Then, it iteratively removes the highest betweenness centrality edge, if by removing it a new community of the desired size is created.
Removal of edges continues until the specified number of communities is met.
Users from a community are then mapped on the same peer.
However, as previously mentioned, neighboring communities in the social graph were mapped on random peers worldwide.
This setup forced both geographically-close and socially-connected communities to be stored on random peers across the globe, enabling us to examine worse case random network delays.

\subsubsection{Synthetic Workloads}\label{emulators_setup}

We emulated the workload of a social sensor and two social applications based on previous system characterizations.
The sensor tested the platform's ability to manage and incorporate new social input under high-stress load.
The user applications tested the end-to-end performance of the \emph{neighborhood} and \emph{social\_strength} requests.

{\it Edge weight update:}
We emulated a Facebook social sensor using the probability for users to post comments on walls and photos extracted from~\cite{wilson09facebook}.
To emulate an interaction from $ego$ to $alter$, a random $ego$ was selected based on their degree, and a random $alter$ from $ego$'s 1-hop friends.
The weight of each input was kept constant to $0.01$ for all users.
Since users were picked based on their degree, users with higher degree probabilistically produced more input, leading to higher weights on their corresponding edges.

{\it Workload for \textit{Neighborhood} Inference:}
A neighborhood request is a limited-distance flood in the network, similar to a tweet in Twitter.
We used a Twitter trace analysis~\cite{krishnamurthy08Twitter} to associate a tweet with a neighborhood request (centered at the source of the tweet) in Prometheus.
Users submitted neighborhood requests probabilistically based on their Twitter degree.
The number of hops for the request was randomly picked from $1$--$3$ hops and the weight was randomly picked, with an upper bound of $0.1$ to produce maximum request load in the system.

{\it Workload for \textit{Social\_Strength} Inference:}
We used an analysis of BitTorrent traces to emulate the workload of a battery-aware BitTorrent application~\cite{king09battorent} on mobile devices: a user may rely on social incentives to be allowed to temporarily ``free ride'' the system when low on battery.
Members of the same swarm check their social strength with the needy leecher to see if they want to contribute by uploading on her behalf.
We assumed that users participated at random in BitTorrent swarms.
Two users were randomly selected as the source and destination of the social strength inference request.
The source user was associated with a total number of requests she would submit in the experiment.
This number was extracted from a BitTorrent trace analysis~\cite{lei05Bittorrent}.

\subsubsection{End-to-end performance on PlanetLab}

\begin{figure*}[tbph]
	\centering
	\hspace{-3mm}
	\subfigure[10 users per peer]{ 
	\hspace{-6px}
	\includegraphics[scale=0.5]{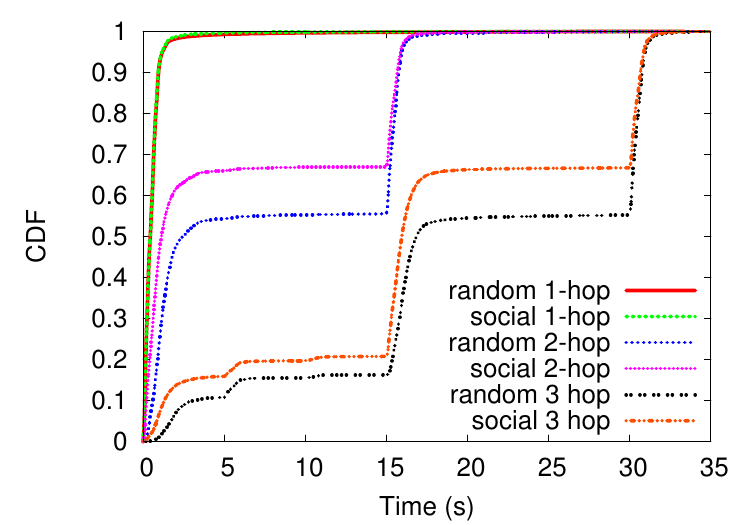}}
	\subfigure[30 users per peer]{
	\hspace{-12px}
	\includegraphics[scale=0.5]{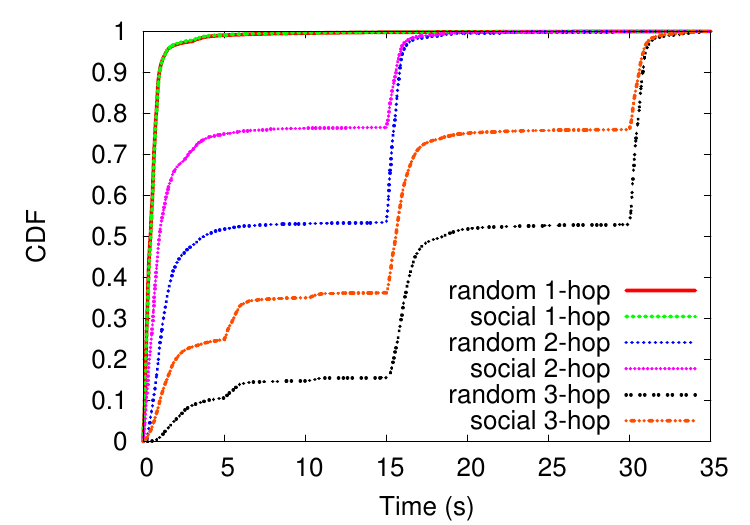}} 
	\subfigure[50 users per peer]{
	\hspace{-12px}
	\includegraphics[scale=0.5]{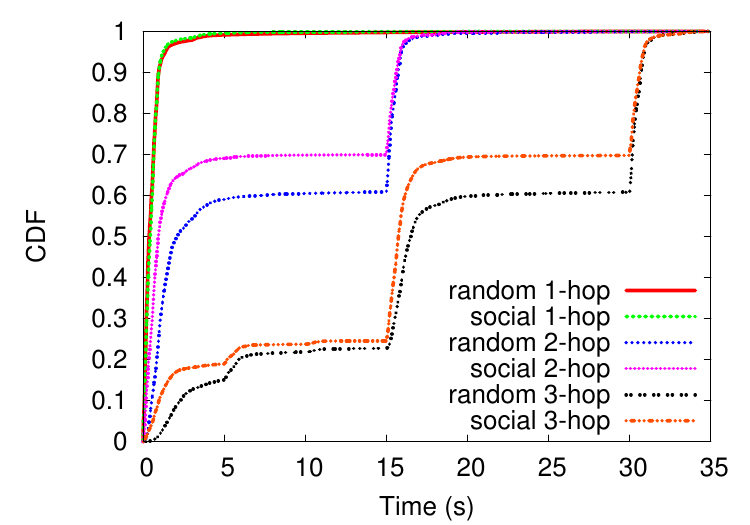}}
	\subfigure[3-hop timeout]{
	\hspace{-12px}
	\includegraphics[scale=0.5]{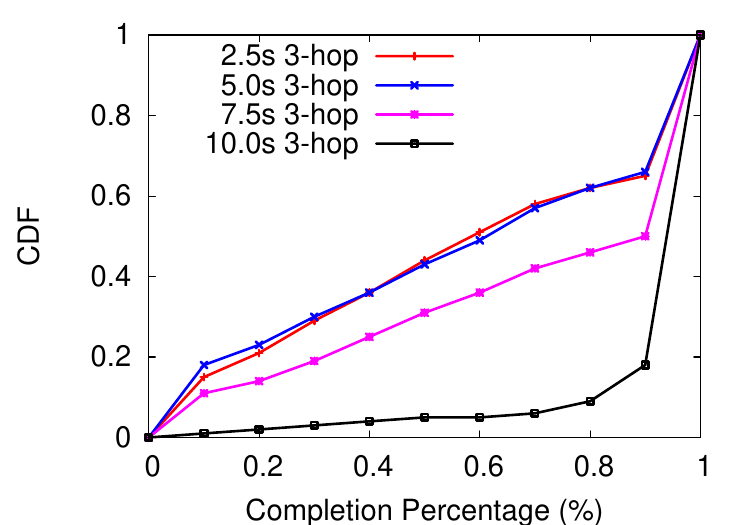}}
	\hspace{-2mm}
	\caption{
	PlanetLab experiments on the $1k$-user graph. (a)--(c) CDF of average end-to-end response time of 1-3 hop neighborhood requests for social (S) vs. random (R) mapping of users on peers, and $10$-$50$ users/peer.
	(d) CDF of average completion percentage of $3$--hop neighborhood requests for varying timeouts.}
\vspace{-3mm}
\label{fig:neighborhood-timeout}
\end{figure*}

For every setup, more than 1 million \textit{social\_strength} and \textit{neighborhood} requests and more than 100 thousand edge updates were submitted from the emulated applications and social sensors running on the $PL$ peers.
Figures~\ref{fig:neighborhood-timeout}$(a)$--$(c)$ show the CDF of end-to-end average response time for the {\it neighborhood} inferences and the different setups.
Since the {\it social\_strength} inference checks all possible 2-hop paths between two users, its results are almost identical to {\it 2-hop neighborhood} inferences and are omitted for brevity.

As expected, the measured response time was higher than the simulations due to overloaded testbed: high average RTT ($125$ms), overloaded CPUs and network interfaces cards (it took 1-2 seconds just to establish a reliable $TCP$ connection for the direct peer communications).
Furthermore, the response time greatly depended on the hops requested.
Larger hops lead to more peers contacted per hop (e.g., for a 3-hop request, and for $10$ users/peer in social (random) mapping, $\sim$$37$($48$) peers were contacted and $\sim$$350$ users returned).
Thus, though the result aggregation is executed in parallel per hop, multiple peers must be contacted, which leads to dependency on a few slow-responding peers.
We observe this in the flat regions of the CDFs, where $\sim$$10$ of $15$secs timeout were spent waiting.
However, an application can reduce the timeout to receive earlier some (incomplete) results (shown next).

\subsubsection{Response Time vs. Completion Rate}

The previous experiments confirmed that a longer timeout (e.g., $T$=$15$seconds per hop) offers a high completion rate, even though the majority of this timeout ($\sim$$2/3$) could be spent waiting for just a few slow peers to respond.
We designed a second set of experiments to measure the tradeoff between response time and response completion rate.
We varied the timeout $T$ to $2.5,5.0,7.5,10.0$ seconds and used the social mapping with $30$ users/peer.
Figure~\ref{fig:neighborhood-timeout}$(d)$ shows the CDF of the average completion percentage of the $3$-hop neighborhood requests under different application-set timeouts.
The results for $1$ and $2$ hops showed very high completion for all requests and timeouts and are omitted for brevity.
Overall, we observe a clear tradeoff between request completion rate and application waiting time for response.
For example, when $T$=$7.5(10)$seconds, about $50\%(80\%)$ of requests have more than $90\%$ completion.
Thus, the longer an application is willing to wait, the more complete the information returned by the social inference is.
A real-time social application---e.g., ``using the \emph{proximity} inference, invite my $2$-hop football contacts for celebration of the team's victory''---could set a low timeout for quick, yet incomplete, results vs. the GamePartnerFinder application that can wait more time to get complete results.

\subsection{Summary of lessons}

From the above simulation and PlanetLab experimental results we draw the following lessons on Prometheus' performance.

\textbf{Lesson 1:}
\emph{The social-based mapping of user data on peers improved the response times by $20$-$25\%$ and reduced the message overhead by $40$--$65\%$ in comparison to random.}
Such gains are observed in the cases of $10$-$30$ users/peer (less for $50$ users/peer), and for $2$-$3$ hops; the $1$-hop is computed either locally at the submitting peer or the first available trusted peer of the source user.
Increasing the graph size and consequently the P2P network leads to higher overhead, but the social-based system exhibits half the overhead of random, both in simulations and in real deployment.

\textbf{Lesson 2:}
\emph{Increasing the service availability (through data replication) and number of users/peer does not always improve the response time and message overhead.}
In general, since requests for a user can be fulfilled by any of her trusted peers, we observe that increasing the data availability by a factor of $3$, and correspondingly increasing the number of users per peer, improves the end-to-end performance by up to $25\%$ and reduces the message overhead by $\sim$$30\%$.
This performance gain is due to two reasons.
First, having more user data on each peer allows more requests to complete with fewer network hops.
Second, given peer churn and vulnerable P2P communication, more service availability per user means more alternative trusted peers to contact for a request to be fulfilled faster.
However, when applying $K$=$5$ and $N$=$50$, we observe reduced improvements due to increased workload on each $PL$ peer from the additional users mapped.

\textbf{Lesson 3:}
\emph{Caching inference results and geographic data decentralization can improve scalability and response time.}
Our simulations showed that caching leads to $20$-$40\%$ message overhead reduction for request execution on larger systems with hundreds of thousands of peers and effectively allows the system to scale easier to larger graphs.
Also, in our real deployment, by placing socially-close communities on random peers, we examined a pessimistic experimental scenario of longer than expected delays for request execution.
However, in reality, we expect neighboring communities placed in geographically-close peers (e.g., same country) instead of random.
In fact, in our simulations we observed that expected network times in a cached, social-based system with in-country delays, were $4$ times smaller than in a non-cached social-based system with random delays.
Given the similar network delay distributions observed in the $PL$-trace and our deployment,\footnote{the $PL$-trace has an average RTT for random delays $4$ times that of in-country delays over $21$ countries, and the average RTT in our deployment for random delays was $6$ times that of in-country delays over $35$ countries.} we conjecture that geographic data placement of social groups on peers, in combination with caching of results, can lead to multiple times shorter response times.
Given that social graphs rarely change~\cite{golbeck07osnchange}, background pre-computation and caching of results should allow the inference execution to scale easier to hundreds of thousands of peers, as shown in our simulations.

\section{Mobile Application Performance}\label{sec:callcensor}

We validated the usability of Prometheus as a social data management service by developing a mobile social application, CallCensor, that utilizes the Prometheus inference functions through the exposed API, under real-time constraints.
\tc{Past work (ContextPhone~\cite{raento05contextphone}) described the ContextContact application which offers cues to the caller about the callee's social context such as location, collocation with other people, phone ringer status, etc.
Our application builds on these lessons and allows the callee's phone to adjust the phone ring based on the owner's social context and the social relationship with the caller.}
We measured its end-to-end performance using a real multi-graph of $100$ users.

The CallCensor application leverages social information received from Prometheus to decide whether or not to allow incoming calls to go through.
For each incoming call, the application queries Prometheus with a {\it social\_strength} or {\it neighborhood} inference request to assess the type of social connection between the caller and the phone owner.
Based on the owner settings (e.g., don't allow personal calls while at work), the application decides if the phone should ring, vibrate or silence upon receiving the call.
The application was written in Java for devices running Google Android OS and was tested on a Nexus One mobile phone from HTC (1GHz CPU, 512MB RAM).

There are multiple scenarios a caller can be connected to the phone's owner; we tested three: directly connected within $1$ social hop, indirectly connect by $2$ social hops, and connected with a high social strength.
We tested each of these scenarios $50$ times.
For each of them, the \emph{ego} and \emph{alter} were randomly chosen, and the inference request was sent to a random peer.
\tc{We assumed users defined ACPs allowing access to all their data, thus enabling requests to proceed over multiple hops, and consequently stressing the system at maximum possible load.}
We measured the end-to-end response time of an inference request submitted to Prometheus. 
This experiment introduced additional overhead due to the communication between the mobile application and Prometheus, and the processing time by the mobile application.

\subsection{Social Multi-Graph from Real Traces}

The social graph used in the CallCensor application experiments was based on data collected at NJIT~\cite{pan11njit-graph}.
The graph has two types of edges, representing Facebook friends and Bluetooth collocation.
Mobile phones were distributed to students and collocation data (determined via Bluetooth addresses discovered periodically by each mobile device) were sent to a server.
The same set of subjects installed a Facebook application to provide their friend lists and participate in a survey.
The user set was small ($100$ users) compared to the size of the student body ($9,000$), therefore resulting in a somewhat sparse graph.
The collocation data have two thresholds of $45$ and $90$ minutes for users to have spent together.

\begin{figure}
	\centering
	\includegraphics[scale=0.2]{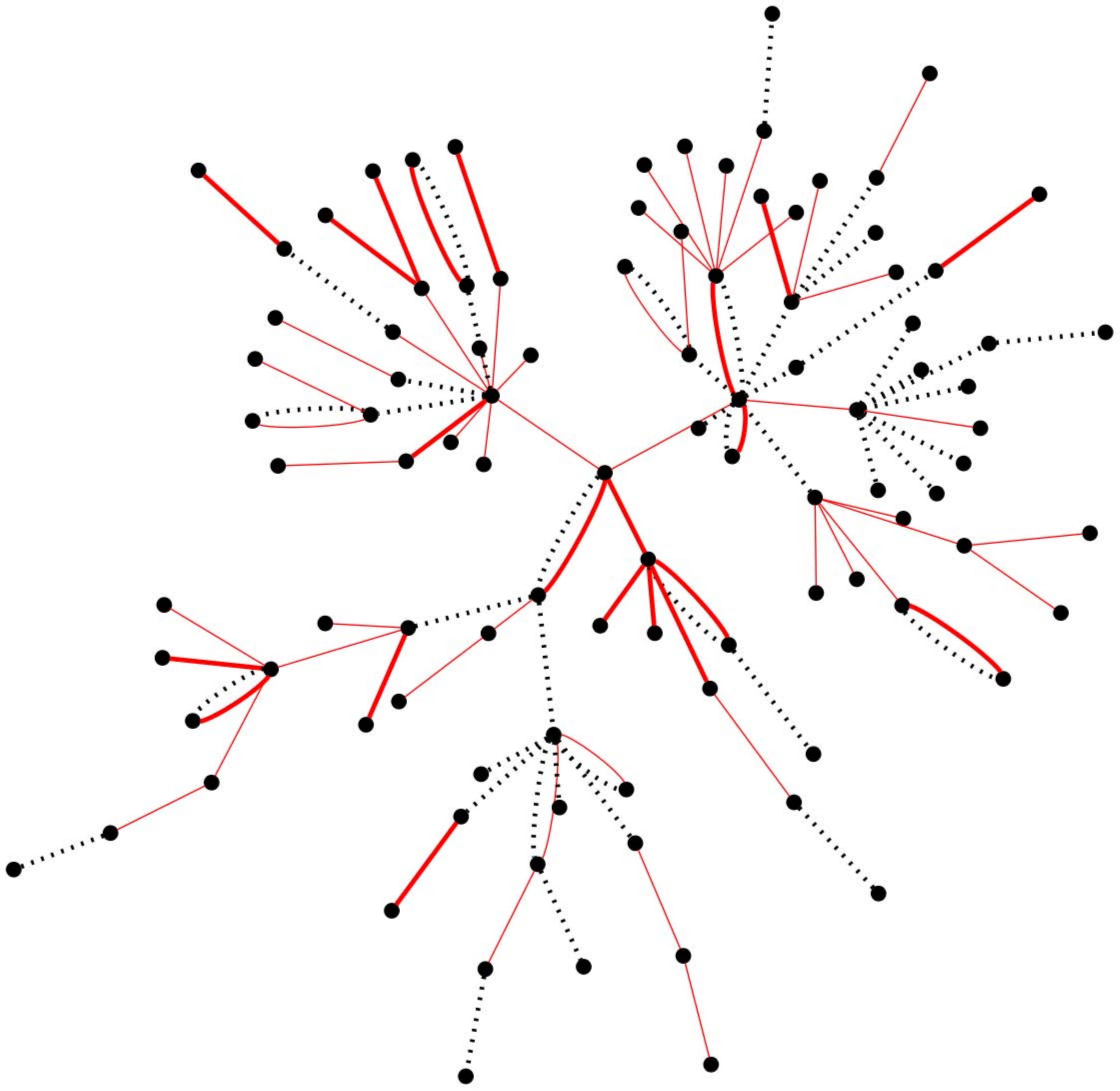}
	\caption{
		Real multi-graph with Facebook edges (black dashed lines) and collocation edges (red continuous lines).
		Line thickness demonstrates edge weight.}
\label{fig:njit}
\end{figure}

While the graph edges were not initially weighted, we applied synthetic weights of $0.1$ for ``Facebook'' edges, $0.1$ for ``collocation'' of $45$ minutes and $0.2$ for ``collocation'' of $90$ minutes.
For the mobile application experiments, we consider the ``collocation'' edges to represent a work relationship, while the ``Facebook'' edges represent a personal relationship.
The user (\emph{ego}) was assumed to be in a work environment when another user (\emph{alter}) called.
\tc{As shown in Figure~\ref{fig:njit}, the multi-graph provided for better connectivity between users since neither the ``Facebook'' nor the ``collocation'' graph is connected, but the graph containing both types of edges is.
We equally distributed the $NJIT$ graph with a social mapping on three PL nodes.
Splitting this $100$ user-graph on $3$ PL peers implies a similar ratio of users/peer as before, while testing $1$ and $2$ hop inference execution on a real multi-graph.}

\subsection{Experimental Results}

\begin{figure}
	\centering
	\includegraphics[scale=1.2]{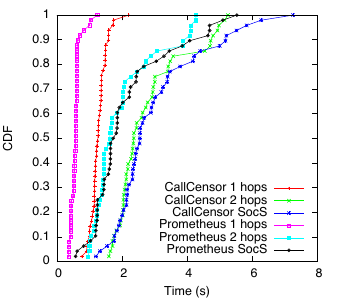}
	\label{fig:callcensor_decentralized}
\caption{
CDF of average end-to-end response time for CallCensor, under three social inference function requests: 1, 2 hops \textit{neighborhood} and \textit{social\_strength (SocS)}.}
\label{fig:CallCensor}
\end{figure}

Figure~\ref{fig:CallCensor} presents the performance for the requests sent by CallCensor, for each of the three scenarios examined.
The results show the time spent by the requests only in Prometheus and the overall time needed by the CallCensor to request and handle a response.
We first observe that the results meet the real-time constraint of the application: the response must arrive before the call is forwarded to the voicemail of the callee (we used the default voicemail time setting).
Second, we notice that the application itself introduced a significant overhead: for example, as much as $100\%$ in the $1$--hop \textit{neighborhood} and $50\%$ in the $2$--hop \textit{neighborhood} and \textit{social\_strength}, due to both communication overhead and execution time on the mobile phone.
Third, we confirm the similarity of the \textit{social\_strength} results with the \textit{neighborhood} for $2$ social hops, as found in the first set of experiments.
\tc{Forth, the $2$-hop request results using this small social graph are similar in performance with the larger $1000$-user graph used before (social mapping, $30$ users/peer).
In particular, more than $80\%$ of requests finished within $5$ seconds in both setups, which leads us to believe that similar performance of the mobile application could be expected in a larger social graph distributed over hundreds of peers.}

\section{Resilience to Malicious Attacks}\label{sec:resilience}

In this section, we discuss how the design characteristics of Prometheus enable the system to defend against attacks
(1)~at the infrastructure level, 
(2)~at the social graph level, 
(3)~at the service level, and
(4)~at the application level, or mitigate their effects on users.

Attacks at the infrastructure level may attempt to install malcode in the system's software or firmware, or to hinder access with a denial of service attack.
Both centralized and decentralized systems are vulnerable to such attacks.
A decentralized system, however, can be the target of additional infrastructure attacks, such as to disrupt system communications (routing of messages between peers) or content storage.
Solutions to eclipse and other infrastructure attacks in DHT-based systems (such as Prometheus) have been discussed in the past~\cite{urdaneta11dht-security}.

Attacks on the social graph aim to manipulate the graph structure by modifying edges, creating new edges with malicious users\footnote{http://www.trendmicro.com/cloud-content/us/pdfs/security-intelligence/white-papers/wp\_the-real-face-of-koobface.pdf} or bias honest users to reciprocate edges to attackers.\footnote{http://ca.olin.edu/2008/realboy/index.html}
A reciprocal edge to an attacker, as implemented in most OSNs, grants him partial or complete access to the victim's data (i.e., online profile and personal information) and can be used for email spam and phishing campaigns.\footnote{http://www.symantec.com/connect/blogs/social-network-attacks-surge}
Centralized systems have complete knowledge of the social graph and monitor its changes to detect such malicious activity (e.g., Facebook Immune System~\cite{stein11fis}).
However, the effectiveness of their defense systems is currently limited against socialbot attacks~\cite{boshmaf11socialbots}.
Prometheus' directed, labeled and weighted social graph is more difficult to manipulate in such attacks because, while it is relatively easy to create an edge from the attacker to the victim, creating a useful reciprocal edge with appropriate label and weight is not in the attacker's control.
In the future, Prometheus could employ techniques such as SybilLimit~\cite{yu08sybillimit} to reduce further the probability to reciprocate edges to sybil attackers.

Attacks at the service level attempt to manipulate inferences on the social graph, such as drop requests or modify results.
Such attacks are more effective on decentralized systems than in centrally controlled systems due to the different level of guarantees that distributed, user-contributed peers can provide.
We experimentally elaborate more on attacks at this level in the next subsections.

Finally, at the application level, malicious services collect and use social data for spamming, stalking, etc.
Users in centralized systems have typically limited or no control on the exposure and use of their data from such 3rd party services, as seen by numerous email spam and phishing campaigns, cyberstalking cases,\footnote{http://www.haltabuse.org/resources/stats/Cumulative2000-2011.pdf} as well as collection of private information from online social aggregators,\footnote{http://www.spokeo.com/} or the NSA.\footnote{http://www.theguardian.com/world/prism}
Instead, Prometheus enables users to control access of applications to their social data via fine-grained ACPs.
In addition, application designers can make use of the Prometheus API to reduce attacks on their users' data.
For example, an email application could filter spam by allowing incoming emails only when the receiver is reciprocating an edge to the sender with appropriate label and weight, and even using knowledge from the receiver's $2$-hop neighborhood to inform the filtering decision~\cite{garriss06re}.

Prometheus is useful to user applications only if it can support an unhindered execution of inference functionalities on users' social data.
In the next subsections we discuss in more detail and present simulation results that show how the socially-aware design of Prometheus increases the system resilience to attacks at the service level which attempt to manipulate inference requests.

\subsection{Attacks at the Service Level}\label{sec:resilience_requests}

As explained in Section~\ref{sec:design}, if a peer does not have the social data necessary to fulfill a request locally, it sends secondary requests to the appropriate peers.
The number of such secondary peers contacted depends on the request type and number of hops requested (which can translate to multiple P2P network hops) and how the social graph is decentralized on peers.
Further, in a system that protects user privacy such as Prometheus, a requestor cannot distinguish how and from which peer any given item entered the result set.
Consequently, for $Alice$ to mount a successful attack at the service level, she would have to control intermediary peers, which have the opportunity to drop incoming requests, modify intermediate results, or change the parameters of secondary requests sent to new peers.
Moreover, if malicious peers collude (e.g., they are all owned or have been compromised by $Alice$), they can increase the magnitude of the attack at the service level.
Eavesdropping is disallowed due to encrypted P2P communication.
Traffic analysis could reveal some characteristics of the network such as central peers that handle multiple requests.
However, due to the chain-type of inference execution and the encrypted messages, an attacker would not be able to infer if they are requests, replies, etc., what type of requests (e..g, neighborhood), or what were the results returned.

We experimentally investigate how the socially-aware design of Prometheus increases its resilience to these attacks by measuring the opportunity of peers to influence (and thus manipulate) results when serving a \emph{neighborhood} inference request.
We define a peer's \emph{influence} on requests as the fraction of requests that the peer serviced over the total number of requests issued in the system.
A peer's influence on a request increases with the number of hops the graph is traversed, since, probabilistically, there are more chances for the peer to participate in the request's execution.
We do not consider the first hop (i.e., the source peer) of a request as malicious, since if it is, no results can be considered legitimate.
In these experiments we consider the worst case scenario where all requests traverse the whole graph: no restriction of the social edge label or weight, all edges are reciprocal, and users define ACPs allowing access to all their data.

We extended our preliminary work~\cite{blackburn11vulnerability} with two sets of experiments to assess the influence of peers in inference execution on real social graphs.
During the experiments, an $n$-hop neighborhood request was performed for all $ego$s (users) in the social graph.
This request was submitted to $ego$'$s$ peer, i.e. peer $P_0$, which could fulfill requests regarding information \emph{only} about users mapped to $P_0$.
For users in subsequent social hops from $ego$ that $P_0$ did not have social data, $P_0$ found the peers storing the particular users' data and submitted secondary requests to be fulfilled by those peers.
Each time a peer served a secondary request, we increased the peer's \emph{influence}.
Similarly with Section~\ref{sec:sim-exp-setup}, we validated our simulator for the influence measurements and report only statistically significant results.

\subsection{Peer Influence in Independent Peer Attacks}

In our first set of experiments we studied the peer influence on five graphs based on real traces from diverse application domains, such as P2P file sharing ($Gnutella$), email communications ($Enron$), trust on consumer reviews ($Epinions$) and friendships in a news website ($Slashdot$).
The social properties of these networks\footnote{Source: http://snap.stanford.edu/data/} have been studied in the past~\cite{ripeanu02gnutella,gomez08slashdot,leskovec09communitystructure} and represent possible activities that could connect Prometheus users with social edges of different labels.

We considered all networks bidirectional with equally-weighted edges and used only the largest connected component of each graph to ensure reachability between all users.
For social mappings of users to peers for the real graphs, we identified communities with the recursive-based Louvain method used in~\cite{kourtellis11p2pcentrality}, for average community size $N$=$10$, $50$ and $100$, while maintaining the replication factor to one ($K$=$1$).
Due to space, we present results for three of the networks and one of the community sizes but discuss lessons based on all networks and setups; all results can be found in~\cite{kourtellis12thesis}.
We also compare results with the $1k$-user synthetic graph used in Section~\ref{sec:exp-eval-merged}.
Table~\ref{tab:real-nets} presents a summary of the networks and communities found.

\begin{table}[tbh]
\vspace{-4mm}
\centering
\caption{Description of real graphs used (same graph metrics as in Table~\ref{tab:synth-nets}).}
\vspace{-2mm}
\begin{tabular}{rrrrrrrr}
\toprule
Dataset		&	Nodes		&	Edges		&	AD		&	CC		&	ED	&	C		&	ACS	\\
\midrule
Gnutella04	&	\num{10876}	&	\num{39994}	&	7.4		&	0.006	&	5.4	&	1087		&	10.0	\\
Gnutella31	&	\num{62561}	&	\num{147878}	&	4.7		&	0.006	&	6.7	&	6255		&	10.0	\\
Slashdot		&	\num{82168}	&	\num{504230}	&	12.3		&	0.060	&	4.7	&	8206		&	10.0	\\
\bottomrule
\end{tabular}
\label{tab:real-nets}
\vspace{-6mm}
\end{table}

\begin{figure*}[htbp]
	\centering
	\hspace{-5mm}
	\includegraphics[scale=0.6]{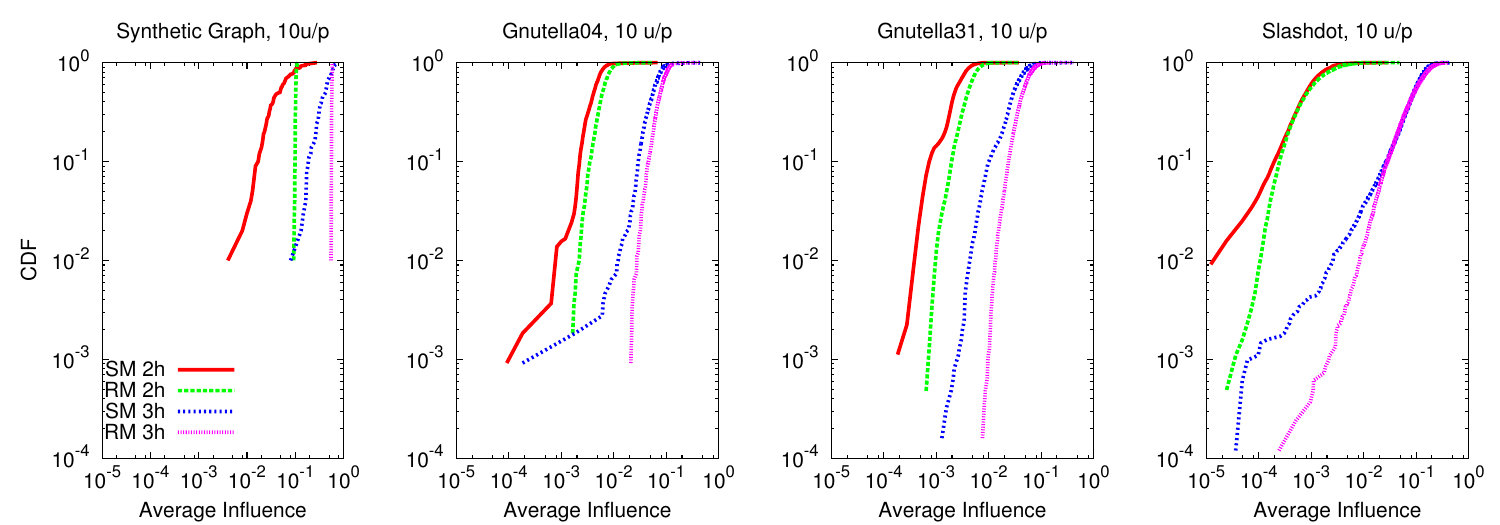}
	\caption{\tc{Peer influence profile: CDF of the average peer influence for the 1000-user synthetic graph, the $Gnutella04$ graph (10.9K users), the $Gnutella31$ graph (62.6K users) and the $Slashdot$ graph (82.2K users), for random ($RM$) and social mapping ($SM$) of 10 users/peer, for $2$ and $3$ hop requests. (Note: axes in log scale).}}
	\label{fig:cdf-combined-resilience-10}
\end{figure*}

\begin{figure*}[htbp]
	\centering
	\hspace{-5mm}
	\includegraphics[scale=0.6]{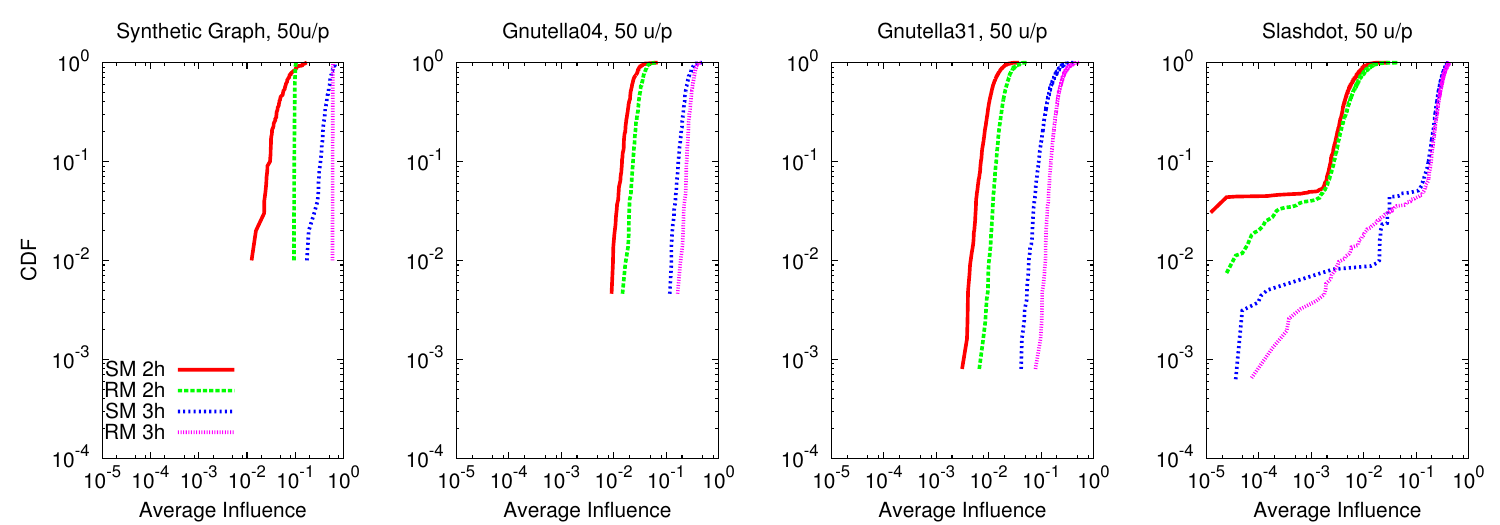}
	\caption{\tc{Peer influence profile: CDF of the average peer influence for the 1000-user synthetic graph, the $Gnutella04$ graph (10.9K users), the $Gnutella31$ graph (62.6K users) and the $Slashdot$ graph (82.2K users), for random ($RM$) and social mapping ($SM$) of 50 users/peer, for $2$ and $3$ hop requests. (Note: axes in log scale).}}
	\label{fig:cdf-combined-resilience-50}
\end{figure*}

Figures~\ref{fig:cdf-combined-resilience-10} and~\ref{fig:cdf-combined-resilience-50} plot the CDF of the average influence of peers in $2$ and $3$ hop requests, for $10$ and $50$ users/peer respectively, for the \textit{Gnutella04}, \textit{Gnutella31} and \textit{Slashdot} graphs and the 1000-user synthetic graph used in Section~\ref{sec:exp-eval-merged}.
The results for all values of $N$ and social graphs can be found in~\cite{kourtellis12thesis}.
From these results we formulate the following lessons.

\textbf{Lesson 4:} \emph{The social mapping reduces the average opportunity of peers to influence requests by $20$--$50\%$ for $2$-hop requests and by $10$--$40\%$ for $3$-hop requests in comparison to the random mapping.}
Figures~\ref{fig:cdf-combined-resilience-10}-~\ref{fig:cdf-combined-resilience-50} show that the random mapping leads to an overall increased opportunity for peers to influence requests, in comparison to the social mapping, regardless of the type and size of graph, and number of hops.
This is because in a social mapping, socially-close users are mapped to the same peers, increasing the likelihood that more hops are served locally when compared to randomly mapped users on peers.
This implies that a social mapping results in fewer secondary requests sent in the system (regardless of the way communities were identified and mapped on peers).
Furthermore, increasing the community size (not shown here) increases a peer's influence on requests, regardless of the type of mapping (random or social), since more requests are served by that peer for each of its users.
However, this influence increase is more prominent in the random than in the social mapping.

\textbf{Lesson 5:} \emph{The peer influence is affected more by the request's number of hops than the type of mapping, topology or the size of the distributed social graph.}
Malicious peers are more effective in small networks, as they can serve and thus influence a larger portion of requests.
For example, \textit{Gnutella04} exhibited a similar influence profile with \textit{Gnutella31} but with higher influence values.
Thus, even though they have the same topological characteristics, in the $6$ times larger \textit{Gnutella31} peers have less opportunity to serve (portions of) requests.
However, if we compare different-hop requests, in our tested graphs and values of users/peer, the average peer influence was higher by $1$--$2$ orders of magnitude in the $3$-hop requests in comparison to $2$-hops.

\textbf{Lesson 6:} \emph{Social-based networks force peers to smaller influence values due to higher clustering coefficient and lower path length.}
\textit{Slashdot} (and similarly \textit{Enron} and \textit{Epinions}~\cite{kourtellis12thesis}), have a different network structure than \textit{Gnutella}, i.e., a higher clustering coefficient that reduces diameter and average path length, but also increases the average degree.
In such social-based networks, the peer influence spreads to a wider range (an order of magnitude smaller values), because queries cover smaller parts of the graph due to higher clustering coefficient and reduced path length.
For these influence values ($<$0.1), the social mapping is more beneficial than random, since it forces requests to touch fewer peers.
However, in higher influence values ($>$0.1), the social mapping performs similarly to random due to the presence of ``hot-spot'' peers.

\textbf{Lesson 7:} \emph{The power-law nature of social graphs allows the formation of ``hot-spot'' peers who serve many requests, thus becoming attractive targets to attacks.}
We observe that in some of the networks, community sizes and $n$-hop requests, the social mapping can lead to a wider distribution of peer influence values, when compared to the random mapping.
This is particularly true in larger networks and community sizes.
However, there are highly influential peers who control more requests flowing through the P2P topology than the average peer \textit{in both types of mappings}.
The emergence of such peers can be attributed to the power-law nature of social networks, where users of high social degree centrality attract a lot of the information flow and requests, regardless of the type of mapping enforced in the system.
Even though such users are closely connected with each other~\cite{shi08well-conn} and are more likely to be mapped together on the same peer in the social than in the random mapping, in our experiments we observed that the random mapping also leads to such highly influential peers, especially in the larger graphs such as $Epinions$ and $Slashdot$.

To tackle the issue of potential attacks, peers can independently monitor the centrality they acquire in the P2P topology based on their users' centrality~\cite{kourtellis11p2pcentrality,kourtellis13socially-aware-designs}.
Consequently, a peer can drop requests from new users to store their social data and encourage them to seek alternative peers, if the new data would create uneven distribution of workload for the particular peer and lead to potential attacks.
Thus, these hot-spot peers can be identified early and targeted for quarantine in the early stages of a malware outburst or other attack.
Alternatively, they can be used to disseminate more efficiently security software patches to handle a malicious attack.

\subsection{Peer Influence in Peer Collusion Attacks}

In the second set of experiments, we investigated the peer influence when peers collude with each other, e.g., they are all controlled by $Alice$.
We assume that $Alice$ attempts to recruit peers for her botnets in two ways.
First, she targets random peers to control different parts of the network; we refer to this as \emph{random collusion} ($RC$).
Second, she targets a cluster of peers that serve a particular portion of the graph, e.g., neighboring groups in a large professional organization; we refer to this as \emph{social collusion} ($SC$), since users mapped on the attacked peers are connected over social edges.

For these experiments we used the largest network \textit{Slashdot} and the social and random mappings with average community size of $10$ users/peer.
We seeded the collusion by selecting $1\%$ random peers (i.e., $1\%$ independent attackers).
Then, we iterated over these peers to expand their collusion sets depending on the collusion type (social or random), until the overall malicious peers (fraction $C$ of all peers) across all collusion sets amounted to a specific portion of the total network.
We varied $C$ in the range of $10\%, 20\%...,50\%$.

\begin{figure}[htbp]
	\centering
	\hspace{-5mm}
	\includegraphics[scale=0.5]{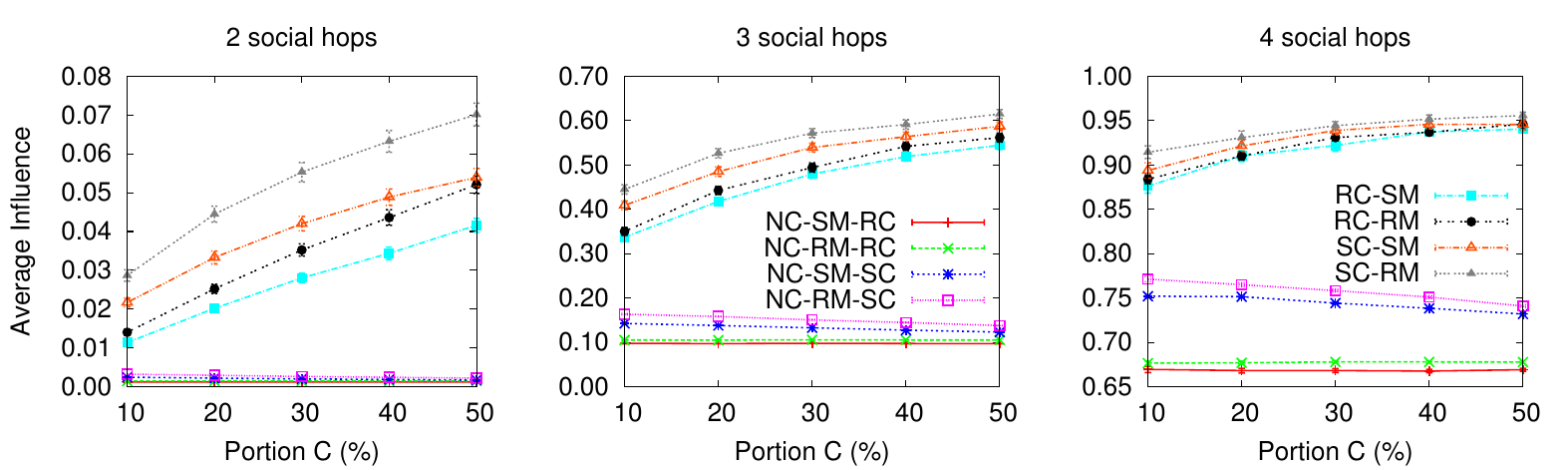}
	\caption{Average peer influence for random ($RM$) and social mapping ($SM$) of $Slashdot$, for $10$ users/peer, $2$--$4$ hop requests and varying range of $social$ ($SC$) or $random$ collusion ($RC$). We compare with the no-collusion ($NC$) scenario.
	Y-bars show $95\%$ confidence intervals.
	}
	\label{fig:SM-RM-all}
\vspace{-4mm}
\end{figure}

Figure~\ref{fig:SM-RM-all} shows the average influence of peers for $10$ repetitions of each of these scenarios.
We observe that the average influence measured on collusion groups ($SC$ or $RC$) is always higher than the average influence of their individual peer members when not colluding with each other ($NC$).
From these results we draw several lessons.

\textbf{Lesson 8:} \emph{Social mappings are more resilient to collusion attacks than random mappings.}
Collusion of peers increases their effectiveness when attacking the system.
However, a random distribution of the social graph onto peers ($RM$) forces requests to access data from more peers than in a social distribution ($SM$), and thus allows peers to control and influence a higher portion of inference requests, either if they are colluding or not.

\textbf{Lesson 9:} \emph{Social peer collusions are less effective on social than random mappings.}
In the social collusion, the attack targets neighboring peers (i.e., their users are directly connected in the social graph).
In this case, the attacker achieves lower peer influence on requests if the graph was mapped with a social than a random mapping.
Overall, the difference between social and random collusion is limited to $\sim$$2\%$ in $2$ hops, $\sim$$10\%$ in $3$ hops and $\sim$$5\%$ in $4$ hops.

\textbf{Lesson 10:} \emph{The peer influence is affected more by the request's number of hops than the collusion size.}
In all tested collusion types and graph mappings, increasing the number of colluding peers increases the average peer influence of an attacker.
However, the peer influence value depends more on the number of hops the request will traverse the graph, than the collusion size: from less than $10\%$ for $2$ hops to more than $90\%$ for $4$ hops (as also shown in the previous experiments).
We also observe highest gains on the rate of change of peer influence for a malicious attacker at $3$ hop requests than $2$ or $4$ hops (the $4$ hop requests cover most of $Slashdot$).

Overall, our experiments demonstrated that the social mapping of user data onto Prometheus peers leads to improved resilience to attacks during request execution, since peers have reduced opportunity to manipulate requests and their results, in comparison to a random mapping (e.g., using a DHT).
Furthermore, we observed that a social peer collusion can be more effective than a random peer collusion.
Therefore, users should carefully select their trusted peers, since unverified peers from their social neighborhood can potentially have increased influence on their requests than random peers.
Finally, there should be default ACPs to control the number of (social and network) hops that requests can travel in the system, since this parameter is more important with respect to peer influence and data exposure, than the extend of peer collusion.

\section{Conclusions}\label{sec:discussion}

We presented Prometheus, a P2P service that provides decentralized, user-controlled, social data management.
Its directed, weighted, labeled and multi-edged social graph offers an aggregate representation of users' social state.
It enables novel socially-aware applications to mine this rich graph via an API that executes social inferences, while enforcing user-defined access control policies.

We evaluated the Prometheus design via large-scale simulations and a prototype deployment on a large distributed testbed with realistic workloads.
Prometheus is designed as an application-oriented platform.
Thus, we tested its end-to-end performance and showed that deadlines set by application requests can be met without significant reduction in the quality of results.
Additionally, we implemented a proof-of-concept mobile social application that utilizes Prometheus functionalities under real-time deadlines.
Further, we investigated the resilience of Prometheus to attacks by malicious users or peers, and established that the socially-aware design of Prometheus constitutes a resilient P2P system that can withstand malicious peer attacks more effectively than solutions that randomly distribute users' data onto peers.

\section*{Acknowledgments}

This research was supported by the National Science Foundation under Grants No. CNS 0952420, CNS 0831785 and CNS 0831753.
Any opinions, findings, and conclusions or recommendations expressed in this material are those of the authors and do not necessarily reflect the views of the sponsors.\\
The authors would like to acknowledge the contributions of Joshua Finnis and Paul Anderson in the Prometheus code and PlanetLab experiments.

\bibliographystyle{ACM-Reference-Format-Journals}
\bibliography{references-short}

\end{document}